\documentclass{article}

\usepackage[utf8]{inputenc}
\usepackage[margin=30mm]{geometry}
\usepackage{parskip}
\usepackage{hyperref}
\usepackage{url}
\usepackage[pdftex]{graphicx}
\usepackage{textgreek}
\usepackage{longtable}
\usepackage{booktabs}
\usepackage[square,numbers,compress]{natbib}
\usepackage{amsmath}
\usepackage{mathtools}
\usepackage{makecell}
\usepackage{arydshln}
\usepackage{pifont}
\usepackage{xcolor}
\usepackage{tabularray}
\usepackage{multirow}

\usepackage{placeins} 
\usepackage{subcaption} 

\usepackage[labelsep=space]{caption}
\renewcommand{\arraystretch}{1.2}

\newcommand{\revision}[1]{#1}

\title{Training a force field for proteins and small molecules from scratch}

\author{
Alexandre Blanco-González\thanks{Equal contribution},
Thea K Schulze*,
Evianne Rovers,
Joe G Greener\thanks{Email: jgreener@mrclmb.ac.uk} \\ \\
{\normalsize \textit{Medical Research Council Laboratory of Molecular Biology}} \\
{\normalsize \textit{Cambridge CB2 0QH, United Kingdom}}
}

\date{}

\begin{document}

\maketitle

\begin{abstract}
\normalsize
\noindent Force fields for molecular dynamics are usually developed manually, limiting their transferability and making systematic exploration of functional forms challenging.
We developed a graph neural network that assigns all force field parameters for diverse molecules using continuous atom typing.
The freely-available model, called Garnet, was trained on quantum mechanical, condensed phase and protein nuclear magnetic resonance data without the use of existing parameters.
The resulting force field shows comparable performance to current force fields on small molecules, folded proteins, protein complexes and disordered proteins.
It shows similar results to popular approaches for relative binding free energy predictions across a range of targets.
Assessing different functional forms shows that the double exponential potential is a flexible and accurate alternative to the Lennard-Jones potential.
Garnet provides a platform for automated, reproducible force field discovery that brings the benefits of machine learning to classical force fields.
\end{abstract}

\section*{Introduction}

Molecular dynamics (MD) has proven itself useful for understanding biology at the atomic level, but the accuracy of a MD simulation is highly dependent on the force field used \cite{Riniker2018}.
For decades, force fields have been tuned manually to fit reference data from quantum mechanics (QM) and experiments \cite{Case2025, Brooks2009}.
This focus on careful optimisation to specific systems is likely the reason for a number of drawbacks of these force fields.
Firstly, the absence of accurate ``universal'' force fields that use the same parameterisation scheme for all molecule types of interest\revision{, the subject of efforts by the Open Force Field Initiative \cite{Wang2024}}.
Secondly, poor performance on molecule types such as intrinsically disordered proteins (IDPs) \cite{Robustelli2018, Huang2017} or amyloids that are studied less than globular proteins.
Thirdly, the scarcity of reproducible training pipelines that can be run without human intervention \cite{Wang2017, Boothroyd2023}.
The solution, in the machine learning age, is to develop an automated approach that uses data on a variety of systems to train an accurate, transferable force field \cite{vanderSpoel2021, Rocken2024, Frohlking2020}.
This will also help improve force fields in future: once training is automated, various components of the force field can be modified to look for improvements.

In particular, the functional form of molecular mechanics (MM) force fields has not changed in decades \cite{Wang2025}.
This is partly due to the difficulty of generating new parameters from scratch, and partly because the conventional form has proved sufficiently accurate for the simulation of proteins on available time scales \cite{LindorffLarsen2011}.
An important use case currently is binding free energy (BFE) predictions, now used routinely in drug discovery \cite{Wang2015, Baumann2026}.
In this case, subtle interactions between the target and the ligand have a large effect on accuracy and the conventional functional form may be a limiting factor.
The transferability of the force field is also important, as typically both proteins and small molecules are present.
\revision{It is common to combine separate but compatible protein and small molecule force fields for BFE calculations.}

A key step in developing a force field applicable to all molecules relevant in biology - including proteins, nucleic acids, carbohydrates, lipids, small molecules and metals - is to move away from human-assigned atom types.
Recent work has shown that graph neural networks (GNNs) applied to the bonding topology of a molecule can predict most force field parameters when trained on a large dataset of high-accuracy QM data \cite{Takaba2024, Wang2022, Thurlemann2023, Seute2025, Bonanni2025, Chen2024, Zhang2022, Zheng2025, Zheng2026}.
In particular, Espaloma has comparable accuracy to existing force fields and can be used for BFE predictions \cite{Takaba2024, Wang2022}.
\revision{Similar approaches include Grappa \cite{Seute2025} and ByteFF \cite{Zheng2025}.}
These methods are not able to train the Lennard-Jones parameters, however.
This means that they reuse some parameters from existing force fields, limiting their automation, their functional form and possibly their accuracy.
They also tend to use existing water models, but co-training the water model could give more accurate parameters.
\revision{More broadly, machine learning interatomic potentials (MLIPs) show good accuracy but are currently too slow to use for routine biomolecular simulation \cite{Eastman2026}.}

Here we present Garnet, a GNN that can predict all MM force field parameters for arbitrary molecules without reusing legacy force field parameters.
Both QM and experimental data are used during training, with the latter incorporated using ensemble gradient techniques \cite{Wang2014, Thaler2021, Norgaard2008}.
We explore different functional forms and find that the double exponential potential \cite{Wu2019, Horton2023} is amenable to training, whereas the Lennard-Jones potential proved challenging.
The resulting force field is evaluated on small molecules, proteins and BFE predictions and shows performance competitive to established force fields, despite little human intervention during training.
On a practical level, having an accurate force field that can simulate a protein with an unusual post-translational modification bound to a drug and cofactors is useful for studying the complex systems found in biology.

\section*{Results}

\subsubsection*{The challenge of training a complete force field}

Dispersion interactions are essential for simulating biomolecules but are weak \cite{Shirts2007, Kolar2011}.
Consider, for example, a simulation of the GB3 protein in water with the Amber14SB/TIP3P force field \cite{Case2025}.
The mean force magnitude on the protein atoms arising from different interactions can be calculated.
For the Lennard-Jones interaction it is 172 kJ mol\textsuperscript{-1} nm\textsuperscript{-1}, compared to 220 kJ mol\textsuperscript{-1} nm\textsuperscript{-1} for electrostatics, 594 kJ mol\textsuperscript{-1} nm\textsuperscript{-1} for harmonic bonds, 665 kJ mol\textsuperscript{-1} nm\textsuperscript{-1} for harmonic angles and 161 kJ mol\textsuperscript{-1} nm\textsuperscript{-1} for torsions.
For the small molecules found in QM datasets used for training, the contribution is even less due to the lack of solvent and periodicity.
Consequently, learning useful Lennard-Jones parameters solely from density function theory (DFT) data is challenging \cite{Hogan2025}, even when dispersion interactions are treated appropriately by the DFT \cite{Klimes2012}; for MM force fields, experimental properties such as enthalpies of vapourisation are typically used alongside DFT data \cite{Boulanger2018}.
Another complication is the high sensitivity of the $\sigma$ parameter (where the twelfth power is used) and the low sensitivity of the $\varepsilon$ parameter (which gives linear scaling) making optimisation challenging.

\subsubsection*{The Garnet force field}

\begin{figure}
  \centering
  \includegraphics[width=0.95\textwidth]{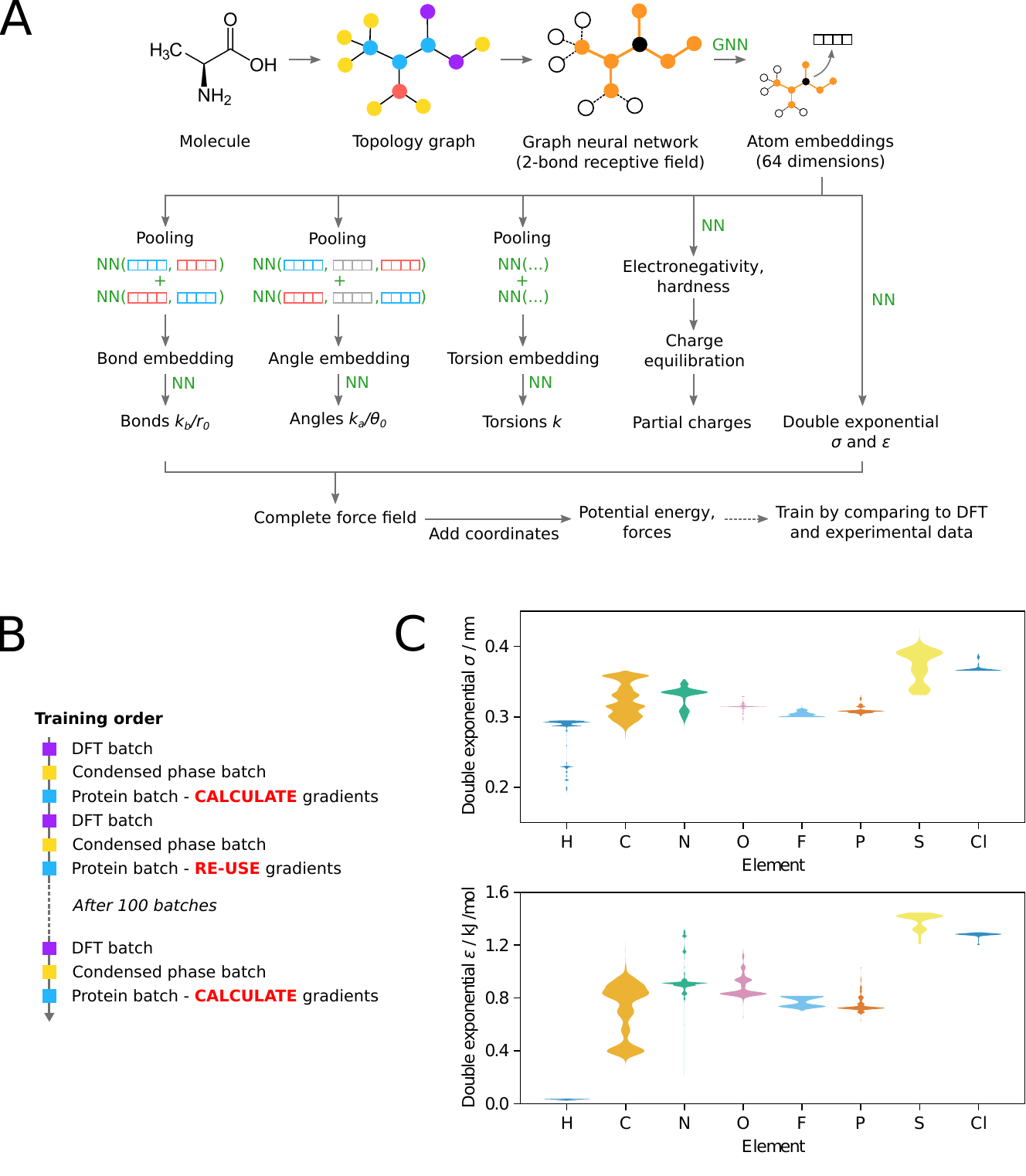}
  \caption{The Garnet model. (A) An overview of the model architecture, which is inspired by Espaloma \cite{Takaba2024, Wang2022}. NN refers to a fully-connected neural network. The double exponential $\alpha$ and $\beta$ global parameters are also trained by the model. (B) The order of batches during training. Calculating gradients to train on GB3 NMR data with ensemble reweighting is slow, so gradients are re-used and new gradients are only computed after 100 batches. A random subset of 200 of the available frames is used to calculate the gradients each time. (C) The double exponential $\sigma$ and $\varepsilon$ parameters for atoms in the SPICE test set calculated with the trained model. $\sigma$ and $\varepsilon$ are similar to the corresponding parameters in the Lennard-Jones potential with $\sigma$ being proportional to the van der Waals radius. By construction, $\sigma$ is allowed to range from 0.05 nm to 0.5 nm and $\varepsilon$ is allowed to range from 0.02 kJ/mol to 1.5 kJ/mol.}
  \label{fig:method}
\end{figure}

To address this problem we trained a model inspired by Espaloma \cite{Takaba2024, Wang2022} but with important differences: enthalpies of vapourisation/mixing and protein nuclear magnetic resonance (NMR) data were targeted alongside DFT data by running simulations during training; different functional forms were explored; DFT forces were split into intramolecular and intermolecular contributions for training; water was not given special treatment and was trained by the model; shallower neural networks were used to reduce the risk of overfitting; and more DFT data was used.
Together, these changes allowed all parameters of the force field to be trained from scratch, though as described in the methods we did use reference trajectories early in training to provide conformations.
This model, called Garnet and shown in Figure~\ref{fig:method}A, was used to generate a force field also called Garnet.

We found that the Lennard-Jones functional form was able to train up to the point where simulations were run during training, for example to fit to enthalpies of vapourisation, but these simulations were not stable.
This is likely due to the high powers in the Lennard-Jones potential making the gradients unstable, or possibly due to the presence of Lennard-Jones interactions for the water hydrogens.
The double exponential potential is an alternative functional form with two global parameters in addition to the $\sigma$ and $\varepsilon$ atom parameters \cite{Wu2019, Horton2023}.
The exponential repulsion, motivated by physics \cite{VanVleet2016}, and the more flexible attractive term could give improved performance for modelling intermolecular interactions.
The potential is only \revision{15-20\%} slower than Lennard-Jones in OpenMM due to the powerful custom forces interface and the speed of exponentials on modern GPUs \cite{Eastman2012, Eastman2017}.
We found that an accurate force field could be trained when replacing the Lennard-Jones potential with the double exponential potential, keeping the other functional forms the same.
Alternative changes to the functional form are discussed later.

The trained double exponential $\sigma$ and $\varepsilon$ parameters are shown by element in Figure~\ref{fig:method}C for the SPICE test set.
The ranges are similar to existing force fields, with a diversity of values shown for carbon depending on its atomic environment.
Hydrogen has low $\varepsilon$ values, indicating that the model does not find it beneficial to involve hydrogen atoms in \revision{these} interactions.
A low-dimensional t-SNE representation of the atom embeddings is shown in Supplementary Figure~\ref{fig:embedding}.
There is a large difference in some hydrogen embeddings, suggesting under-explored discrete atom types for hydrogen in existing force fields.
As shown in Supplementary Figure~\ref{fig:charges}, the Garnet partial charges are over-polarised compared to existing force fields.
This is due to fitting to the MBIS partial charges in the SPICE dataset \cite{Verstraelen2016, Eastman2023, Eastman2024}, \revision{and to the fact that the DFT systems used for training are small and in vacuum, therefore do no capture the appropriate bulk phase charge distributions}.
The parameters assigned to ubiquitin by Garnet and Amber14SB are compared in Supplementary Figure~\ref{fig:parameters}.
\revision{The 1-4 electrostatic scaling factor (0.57) and $\alpha$/$\beta$ global parameters for the double exponential potential (12.2/4.33) are somewhat similar to existing force fields \cite{Horton2023}.}

\subsubsection*{Small molecule benchmark}

First, we assessed Garnet on held out portions of the SPICE DFT dataset used for training as shown in Table~\ref{tab:dft_comparison}.
The error for both forces and potential energy differences is lower than Espaloma and much lower than OpenFF 2.2.1 \cite{Boothroyd2023} across different subsets of the SPICE dataset.
As expected, the error is still much higher than MACE-OFF23 \cite{Kovacs2025}, a slower MLIP.

\begin{table}
  \centering
  \begin{footnotesize}
    \begin{tabular}{ l l l | p{1.2cm} p{1.2cm} p{1.2cm} p{1.2cm} }
      \hline
      \textbf{SPICE subset} & \textbf{n mols} & \textbf{n confs} & \textbf{Garnet} & \textbf{Espaloma \newline 0.4.0} & \textbf{OpenFF \newline 2.2.1} & \textbf{MACE- \newline OFF23} \\
      \hline
                                              &       &        & \multicolumn{4}{c}{\textbf{Force error / kJ mol\textsuperscript{-1} nm\textsuperscript{-1}}} \\
      \hline
      Test Set 1 v1.0                         &   800 &  7,490 & 398 & 522 & 1104 & 40 \\
      DES Monomers Single Points Dataset v1.1 &    15 &    750 & 329 & 452 & 1015 & 14 \\
      Dipeptides Single Points Dataset v1.3   &    16 &    798 & 355 & 488 & 1048 & 35 \\
      Amino Acid Ligand v1.0                  & 2,103 &  4,455 & 269 & 423 &  689 & 27 \\
      DES370K Single Points Dataset v1.0      &    99 &  1,974 & 163 & 371 &  694 & 15 \\
      PubChem Single Points Dataset           &   594 & 29,433 & 500 & 535 & 1173 & 36 \\
      \hline
                                              &       &        & \multicolumn{4}{c}{\textbf{Potential energy diff error / kJ mol\textsuperscript{-1}}} \\
      \hline
      Test Set 1 v1.0                         &   800 &  7,490 & 14.2 & 21.3 & 36.5 & 3.3 \\
      DES Monomers Single Points Dataset v1.1 &    15 &    750 &  4.4 &  4.1 &  9.1 & 0.3 \\
      Dipeptides Single Points Dataset v1.3   &    16 &    798 & 11.8 & 17.8 & 31.5 & 3.3 \\
      Amino Acid Ligand v1.0                  & 2,103 &  4,455 &  6.2 &  5.8 & 16.6 & 2.2 \\
      DES370K Single Points Dataset v1.0      &    99 &  1,974 &  3.5 &  7.9 & 15.0 & 1.2 \\
      PubChem Single Points Dataset           &   594 & 29,433 & 12.6 & 11.1 & 23.2 & 1.8 \\
      \hline
    \end{tabular}
  \end{footnotesize}
  \caption{Performance of Garnet on subsets of the SPICE dataset. For each subset apart from the Test Set, $\sim 3$\% of the total molecules, which were not used for training, are assessed here. For each conformation, the force error is the mean magnitude of the difference between the calculated and DFT force across atoms. The median of these values across all conformations is shown in the table. For the potential energy difference error, pairs of conformations of the same system are considered. The difference in potential energy for the pair is calculated with the model and using the DFT values, and the error is the absolute difference between the model and DFT differences. For MACE-OFF23 the medium model was used. Some conformations gave errors with some of the methods, so only conformations that were successful with all four methods are reported. Conformations containing an atom with a DFT force magnitude greater than $10^4$ kJ mol\textsuperscript{-1} nm\textsuperscript{-1} were skipped. Espaloma and MACE-OFF23 were trained on some subsets of SPICE so will have been trained directly on some conformations used here for testing.}
  \label{tab:dft_comparison}
\end{table}

\begin{figure}
\centering
    \begin{minipage}[c]{0.64\textwidth}
        \begin{subfigure}[c]{\linewidth}
            \includegraphics[width = \textwidth]{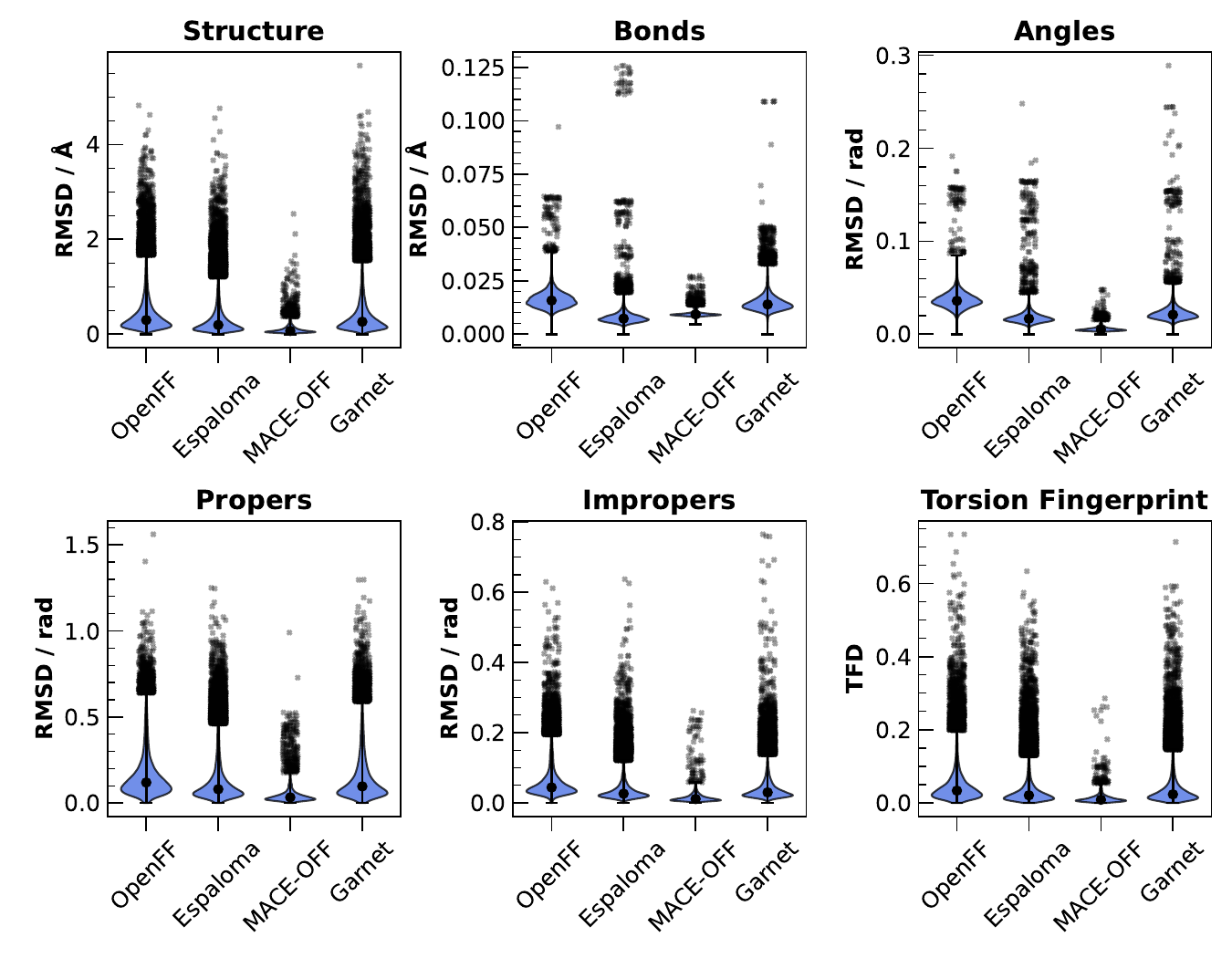}
            \caption{}
            \label{fig:violin_smallmols}
        \end{subfigure}
        \medskip
        \begin{subfigure}[c]{\linewidth}
            \includegraphics[width = \textwidth]{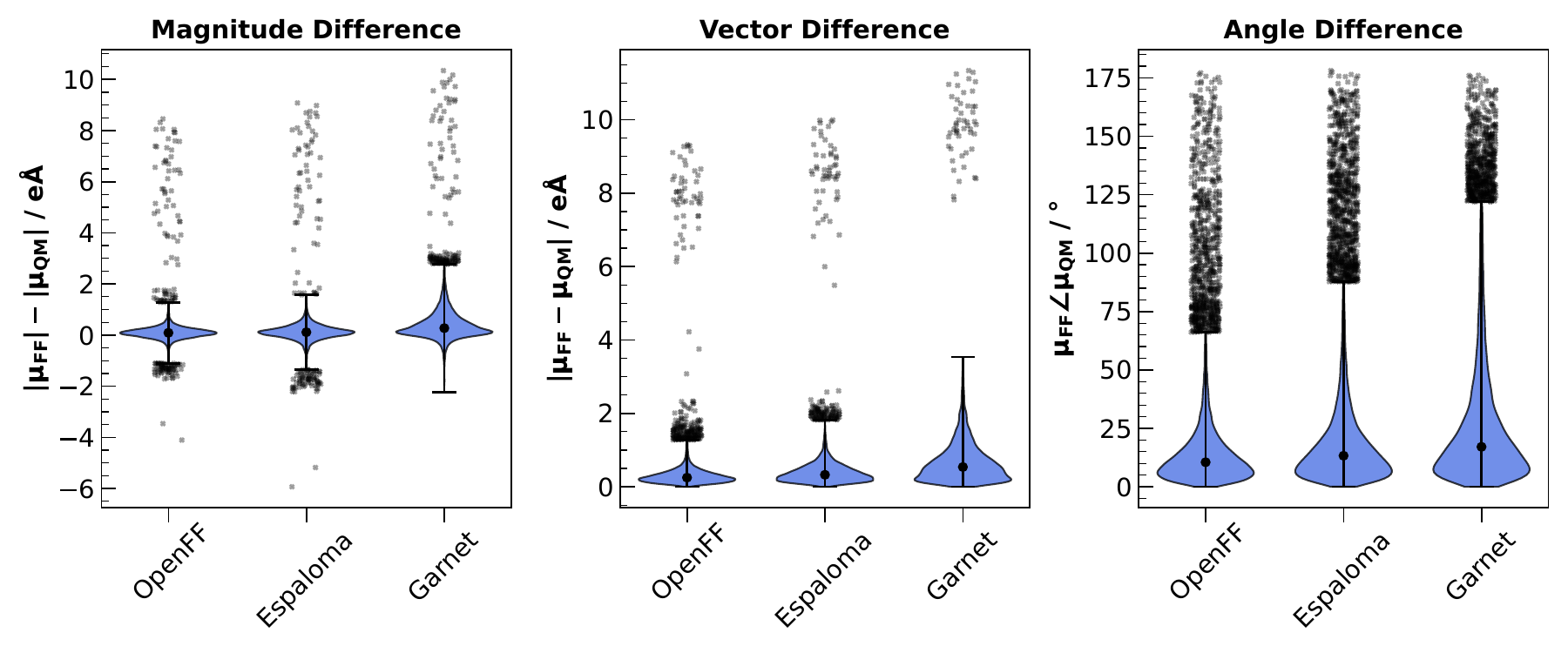}
            \caption{}
            \label{fig:dipoles_vec}
        \end{subfigure}
        
    \end{minipage}
    \hfill
    \begin{minipage}[c]{0.35\textwidth}

        \begin{subfigure}{\linewidth}
            \includegraphics[width = \linewidth]{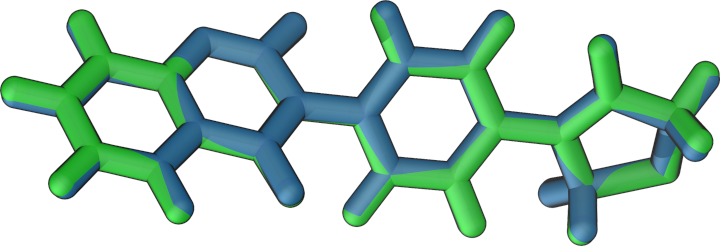}
            \medskip
            \includegraphics[width = \linewidth]{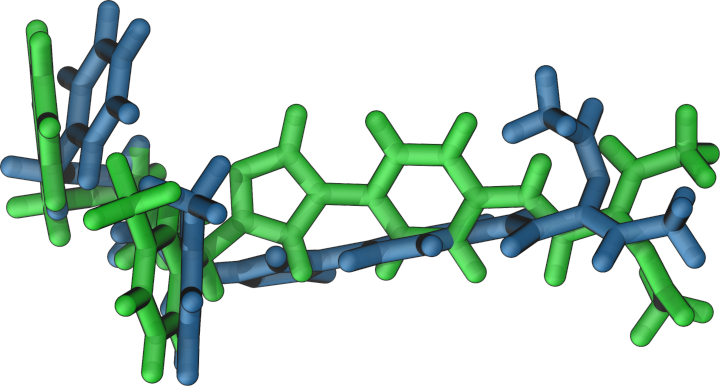}
            \caption{}
            \label{fig:good_bad_fit}
        \end{subfigure}
        \medskip
        \begin{subfigure}{\linewidth}
            \includegraphics[width = \linewidth]{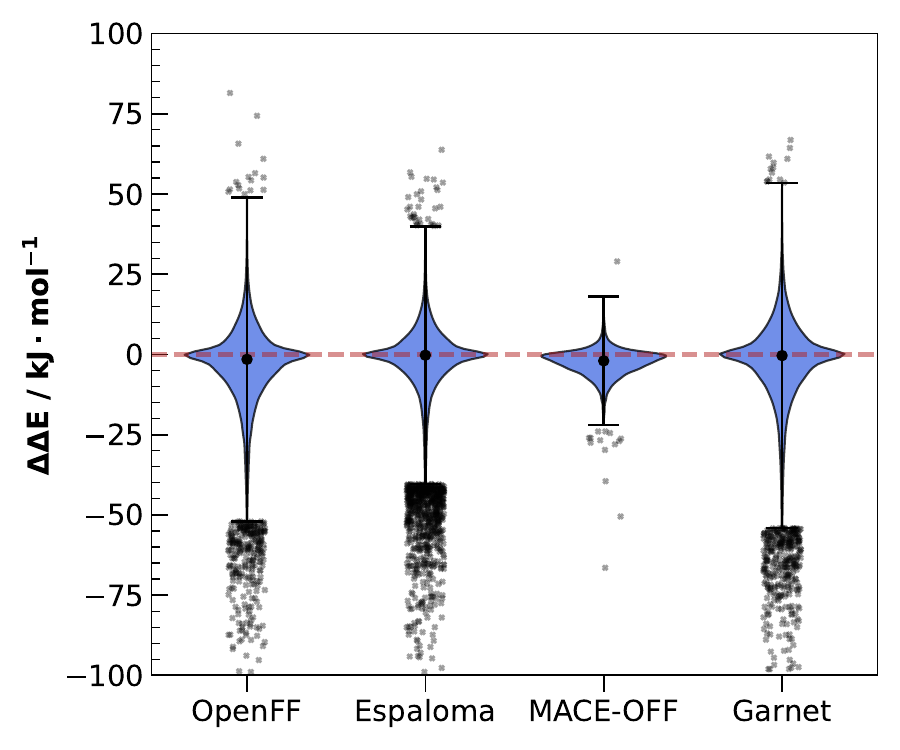}
            \caption{}
            \label{fig:violin_dde}
        \end{subfigure}
        
    \end{minipage}
    \caption{\revision{Structural metrics} of small molecules in the OpenFF Industry Benchmark dataset \cite{Horton2025} and SPICE dataset. (A) Violin plots for the different structural deviation metrics. The whisker range is three times the median absolute deviation, clamped to zero when needed. Points falling outside this range are considered outliers, are not included in the violin density estimation, and are shown as small ``x'' markers. The black circular marker denotes the median of each distribution. \revision{(B) Violin plots for the dipole moment comparison to the QM values}. \revision{(C)} An example of molecules well captured (top) and poorly captured (bottom) by the Garnet force field, with the QM structure in green and the MM structure in blue. \revision{(D)} Violin plots for the energy deviation metric, $\Delta\Delta\mathrm{E}$, showing how well energy differences near the QM minimum are captured by the force fields. \revision{Because $\Delta\Delta\mathrm{E}$ is a signed difference of relative minimized energies, values larger than $k_{\mathrm{B}}T$ may indicate errors in relative Boltzmann weights, but cannot be interpreted directly as population errors.} Values near zero represent accurate energy wells.}
    \label{fig:smallmols}
\end{figure}

Encouraged by this, we measured how well Garnet was able to minimise the conformations of small molecules in the OpenFF Industry Benchmark dataset \cite{Horton2025}. This dataset contains thousands of QM-minimised small molecules, each represented by several conformations. In general, our results are competitive with Espaloma and outperform OpenFF (Figure~\ref{fig:violin_smallmols}). All four methods show distributions that are clearly skewed towards lower values, with long, thin, tails extending towards higher deviations. This indicates that the vast majority of the structures are properly captured.

In contrast to the bond and angle root mean square deviation (RMSD) distributions, the RMSD distributions for proper and improper torsions are clearly wider, indicating that rotational degrees of freedom and the associated energetic barriers are some of the most challenging aspects to capture accurately with MM force fields. The structures of two molecules in the first and last percentiles of the structure RMSD are shown in Figure \ref{fig:good_bad_fit}. The incorrect molecule is poorly captured due to the rotational degrees of freedom, where the rotating bonds have twisted away from the QM conformation. Nevertheless, both Espaloma and Garnet display similar performance for torsions and show a slight improvement over OpenFF. This is further corroborated by the torsion fingerprint deviation (TFD), which provides a torsion-centric measure of conformational similarity \cite{Schulz2012} (Figure~\ref{fig:violin_smallmols}, bottom right).

\revision{
We assessed how well Garnet captures the dipole moments of the held-out test split of the SPICE dataset, and compared this to OpenFF and Espaloma. As described previously, our force field was found to be slightly over-polarised, as seen in Figure \ref{fig:dipoles_vec} and Supplementary Figure \ref{fig:dipole_tensor}. We found that not only the magnitude of the dipole vector was affected, but also its orientation, as per the angle difference subplot. In addition,} an energy deviation metric, $\Delta\Delta\mathrm{E}$, was computed. This evaluates how the relative energies of conformations minimised using different methods differ from their corresponding QM ground truths. All tested classical force fields present more outliers for negative values than for positive ones, indicating a systematic underestimation of conformer energy differences \revision{(Figure~\ref{fig:violin_dde})}. MACE-OFF23 exhibits both fewer outliers and a narrower distribution than its classical counterparts, reflecting better overall agreement with the QM energies. However, it also displays a small negative offset, suggesting that the bias is still present. The three classical force fields demonstrate comparable results: OpenFF shows a slight bias towards negative values, while Espaloma and Garnet are well-centred around zero with Espaloma displaying a slightly narrower distribution.

We analysed chemical motif enrichments for the lowest and highest 1\% structural deviation subsets of the conformers, in order to determine if there are functional groups that are captured better or worse by each force field (Supplementary Table~\ref{tab:compare_structure_best} and \ref{tab:compare_structure_worst}). Here we focus on structure RMSD to describe global conformation. More comprehensive results for the internal coordinate metrics are provided in the data available online.

Garnet performs well compared to the other methods for molecules with the \texttt{imidazole}, \texttt{fluorinated C}, \texttt{ketone} and \texttt{imide} motifs.
By contrast, it does comparatively poorly for the \texttt{alkene}, \texttt{urea}, \texttt{oxazole}, \texttt{conjugated} and \texttt{phosphonate} motifs.
No motif is enriched across all four methods. The most consistent top performers \texttt{pyrazole}, \texttt{aldehyde} and \texttt{thiol} are enriched in three methods. Rigid aromatics and polar groups are enriched in different models.
Conversely, the highest-deviation structures are dominated by complex nitrogen functional groups (\texttt{amide}, \texttt{tertiary amine}, \texttt{sulfonamide} and \texttt{quaternary amine}), failing with at least three methods. This reveals a bias in structure RMSD: it strongly amplifies errors at flexible locations in the molecules.

Comparing these results to internal coordinate metrics exposes some differences. Linear systems (\texttt{alkyne} and \texttt{diazo}) show excellent overall structural agreement despite populating the worst-performing angle RMSD subsets. Because these groups are typically small, or localised at terminal positions, local bending minimally displaces the molecule, allowing structural superposition to alleviate the local failures. Similarly, the hypervalent \texttt{sulfone} motif shows high local structural strain across most models, possibly due to fixed-charge polarization limits.

\revision{Finally, we evaluated Garnet's ability to estimate small molecule condensed phase properties, specifically the density ($\rho$) and enthalpy of vapourisation ($\Delta H_v$). We used a benchmarking set of experimental $\rho$ and $\Delta H_v$ data for 12 small molecules \cite{wang_application_2011, NISTWebBook} containing diverse functional groups with biological relevance, such as hydroxyl, carboxyl, thiol, amine and amide groups.

We first estimated $\Delta H_v$ values from simulations and compared our results to the experimental data. Garnet predicts $\Delta H_v$ with an RMSE of $4.29_{2.12}^{6.12}$ kcal/mol, whereas OpenFF-2.2.1 has an RMSE of $2.63_{0.48}^{4.48}$ kcal/mol. The linear correlation between predicted and experimental $\Delta H_v$ values is above 0.80 for both force fields. We then similarly compared simulated and experimental densities and obtained an RMSE of $0.13_{0.11}^{0.16}$ g/mL for Garnet and $0.041_{0.021}^{0.058}$ g/mL for OpenFF-2.2.1. Both force fields predict $\rho$ with a correlation above 0.9. This is shown in Supplementary Table \ref{tab:density_dhvap_results} and Supplementary Figure \ref{fig:density_dhvap_results}. Garnet (and OpenFF-2.2.1) captures $\rho$ and $\Delta H_v$ better for the three molecules for which $\Delta H_v$ was seen during Garnet training. Garnet tends to overestimate both $\rho$ and $\Delta H_v$, suggesting that the strength of attractive intermolecular interactions are overestimated. Future versions of our model could include $\Delta H_v$ data for more than three small molecules during training, as OpenFF does \cite{Boothroyd2023}. Similarly, we expect that training and benchmarking against small molecule free energies of solvation would be useful.}

\subsubsection*{Protein benchmark}

\begin{figure}
    \centering

    \begin{minipage}[c]{0.4\textwidth}

        \begin{subfigure}{\linewidth}
            \includegraphics[width = \linewidth]{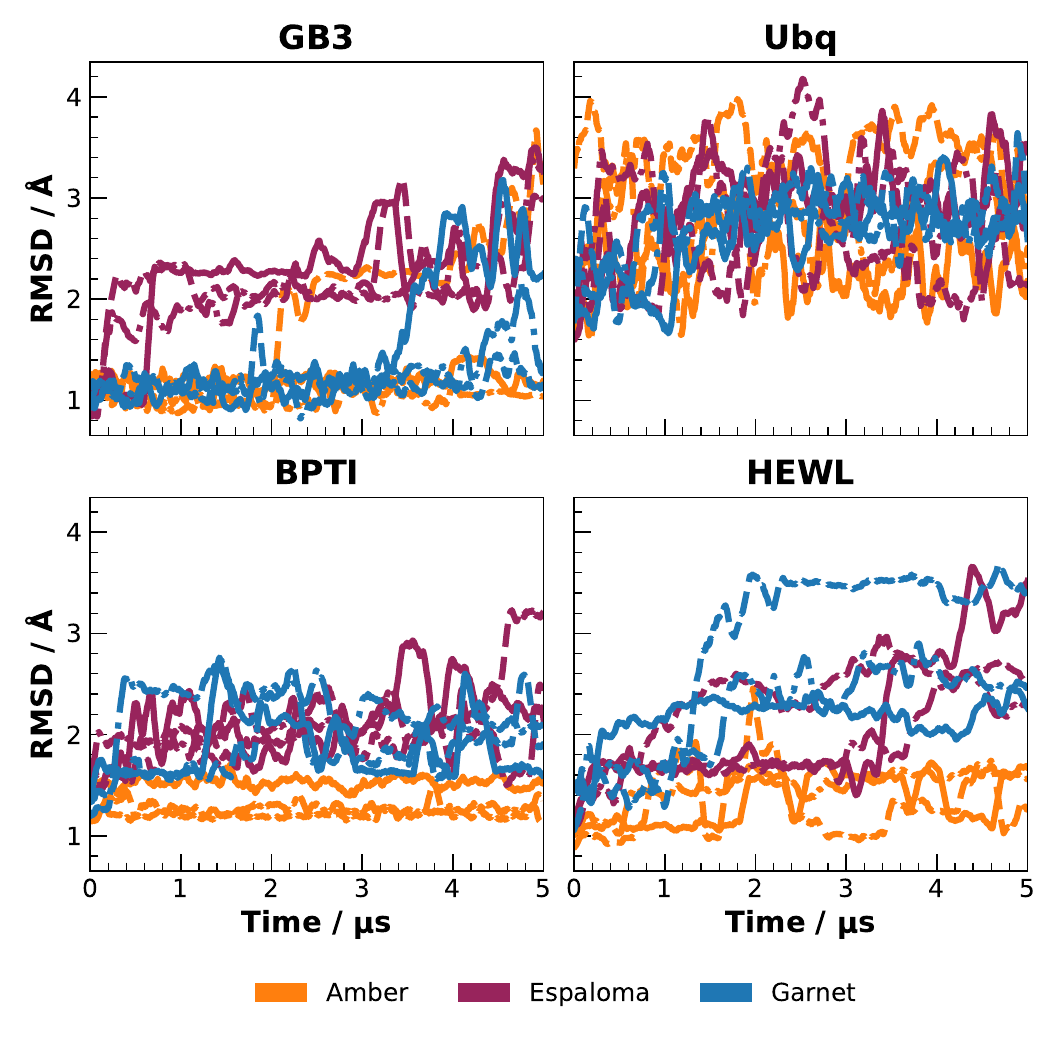}
            \caption{}
            \label{fig:rmsd_val}
        \end{subfigure}  
        \medskip
        \begin{subfigure}{\linewidth}
            \includegraphics[width = \linewidth]{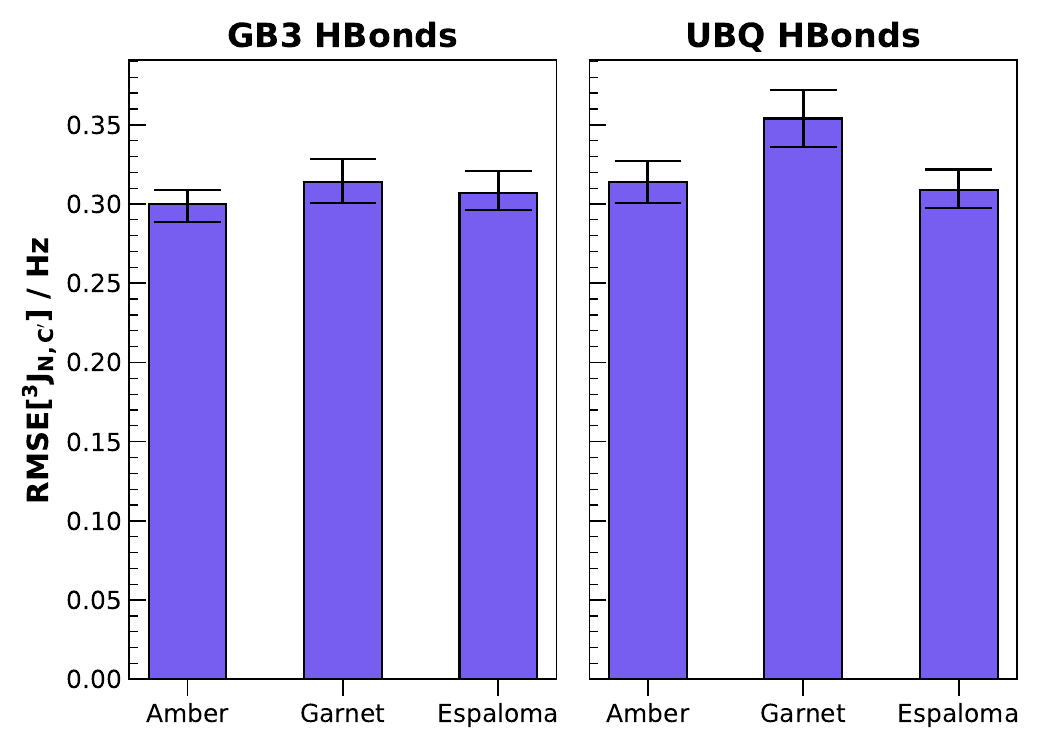}
            \caption{}
            \label{fig:rmse_hbonds}
        \end{subfigure} 
    \end{minipage}
    \hfill
    \begin{subfigure}[c]{0.55\textwidth}
        \includegraphics[width = \textwidth]{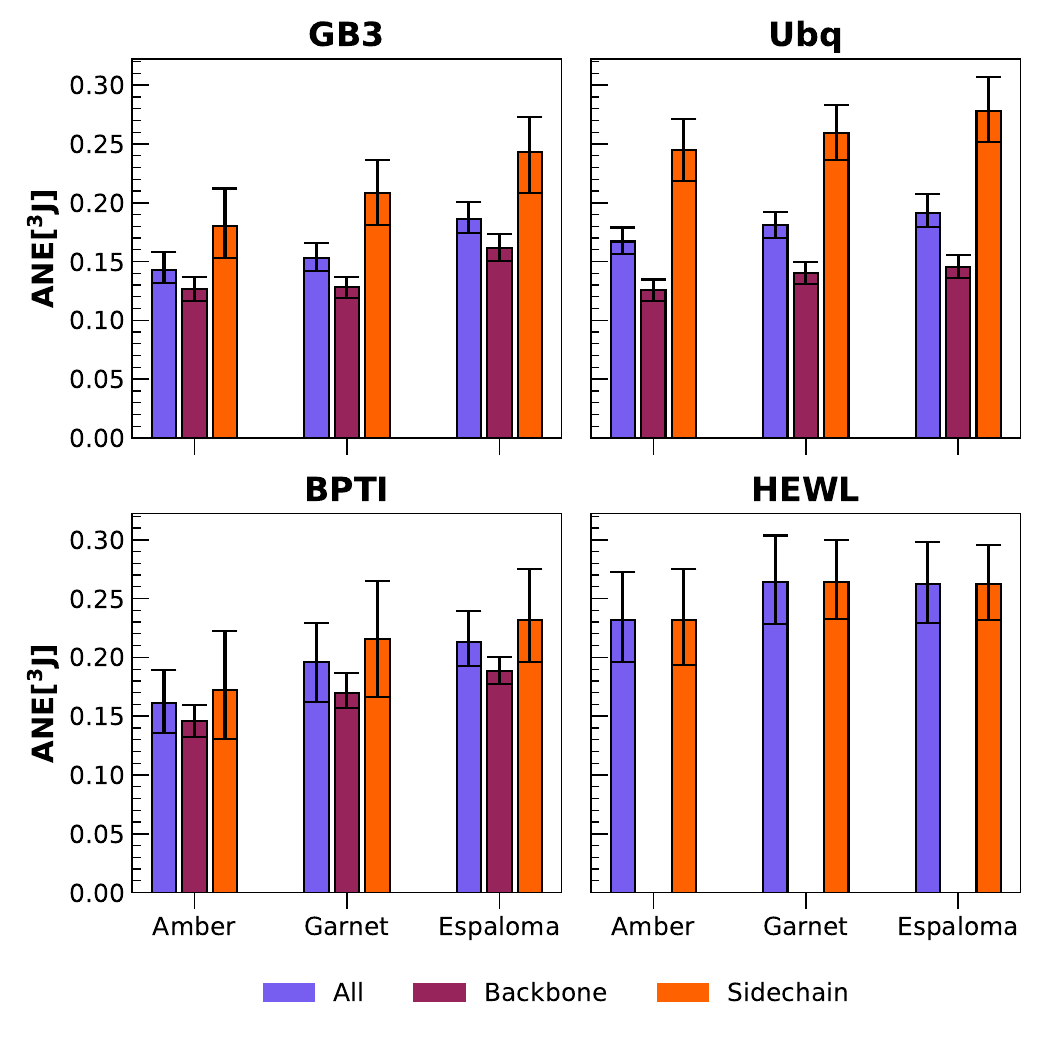}
        \caption{}
        \label{fig:ane_nmr}
    \end{subfigure}
    \caption{Performance of Garnet on simulating folded proteins. (A) RMSD to the native structure for the simulated proteins, each replica represented in a different line style. The RMSD is smoothed by taking the mean over a window of values extending 10 snapshots either side. (B) RMSE for the hydrogen-bond related scalar couplings of GB3 and Ubq. (C) ANE for the scalar couplings of the four studied proteins. In all cases the error bars estimate a 99.9\% confidence interval, computed through bootstrapping. HEWL lacked data for the couplings related to backbone torsions. GB3 was used during model training.}
    \label{fig:nmr_results}
\end{figure}

Next, Garnet was assessed on folded proteins.
Four proteins with available NMR data were stable in simulations of 5 μs: the third IgG-binding domain of protein G (GB3), bovine pancreatic trypsin inhibitor (BPTI), hen egg white lysozyme (HEWL) and ubiquitin (Ubq).
We included GB3, which was used during training, to assess possible overtraining by the model.
The RMSD to the native structure generally remained less than 3 \AA\ across the simulations as shown in Figure~\ref{fig:nmr_results}A.

GB3 and Ubq had available experimental data for scalar couplings arising from hydrogen-bonding patterns associated with the secondary structure of the proteins. Figure~\ref{fig:rmse_hbonds} shows that the root mean square error (RMSE) for the three force fields is similar for GB3. On the other hand, for Ubq Garnet slightly underperforms compared to Amber14SB and Espaloma, exhibiting a larger RMSE by approximately 0.05~Hz.
Considering scalar couplings originating from nuclei connected through three consecutive covalent bonds, all four simulated proteins have available experimental measurements. For every protein, our model performs competitively against both Amber14SB and Espaloma (Figure~\ref{fig:ane_nmr}). In all cases apart from HEWL, Amber14SB shows the lowest absolute normalised error (ANE), followed by our model and then Espaloma. The good performance on GB3 by Garnet on these benchmarks could be due to overfitting. Overall, these results indicate that Garnet is able to reproduce NMR scalar couplings from protein simulations with accuracy comparable to established alternatives.

\begin{figure}
  \centering
  \includegraphics[width=1.0\textwidth]{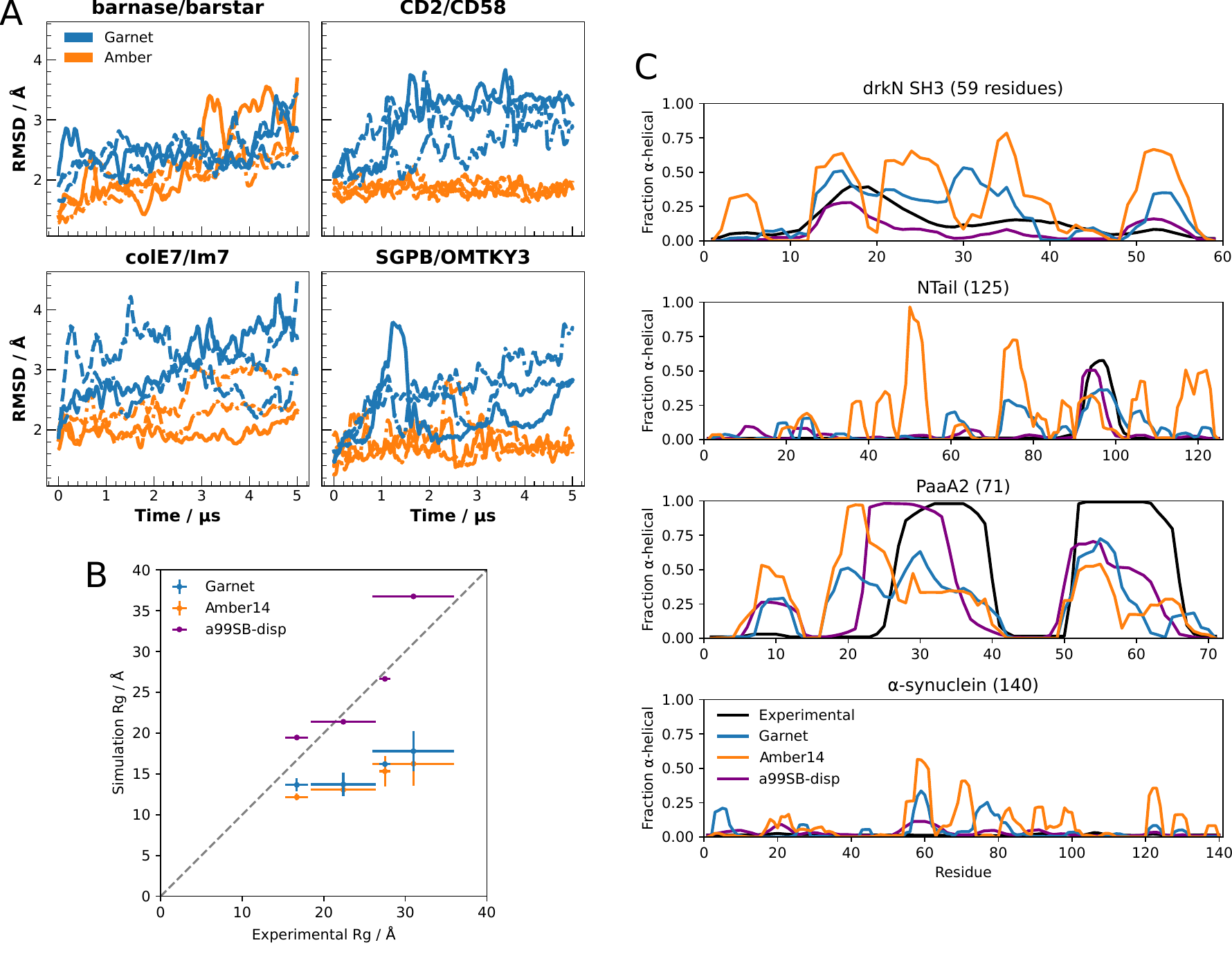}
  \caption{Performance of Garnet on simulating protein complexes and IDPs. Experimental values for $R_g$/\textalpha-helical fraction and simulation $R_g$ values for a99SB-\textit{disp} are taken from Robustelli et al.\ 2018 \cite{Robustelli2018}. (A) Simulations were run for four protein complexes with Garnet or Amber14SB/TIP3P \cite{Case2025}. The RMSD of each complex to its starting conformation is shown over the course of three repeats. The RMSD is smoothed by taking the mean over a window of values extending 10 snapshots either side. The RMSD of the monomers is shown in Supplementary Figure~\ref{fig:complex_monomers}. (B) Comparing simulated $R_g$ to experiment for four IDPs. Each point (excepting a99SB-\textit{disp}) is the mean $R_g$ over 3 simulations of 3 \textmu s, after an initial burn-in period of 1 \textmu s starting from the final frame of a 2 \textmu s simulation in the GB99dms implicit solvent force field \cite{Greener2024}. The error bars in the $x$ direction represent uncertainty in the experimental value and error bars in the $y$ direction represent 95\% confidence intervals of the mean calculated from the standard error of the mean across the 3 simulations. The distribution of $R_g$ values is shown in Supplementary Figure~\ref{fig:idp_density}. (C) The \textalpha-helical fraction for each residue of the four IDPs with different force fields from the same simulations.}
  \label{fig:complex_idp}
\end{figure}

Garnet was also able to keep protein complexes stable, as shown in Figure~\ref{fig:complex_idp}A.
Four protein complexes that are stable on the time scale available for simulation \cite{Piana2020} are generally stable in the force field: barnase/barstar, CD2/CD58, colE7/Im7 and SGPB/OMTKY3.
This indicates that the learned intermolecular interactions are sufficient to keep molecules together.
The corresponding data for the monomers is shown in Supplementary Figure~\ref{fig:complex_monomers}.

Since the GB3 protein used during training is folded, Garnet was not explicitly trained on IDP data.
Nonetheless, we assessed its ability to reproduce the experimental properties of four IDPs \cite{Robustelli2018}: drkN SH3, the N\textsubscript{TAIL} domain of the measles virus nucleoprotein, the ParE2-associated antitoxin PaaA2 and \textalpha-synuclein.
It over-compacts IDPs, as shown by the radii of gyration ($R_g$) in Figure~\ref{fig:complex_idp}B and Supplementary Figure~\ref{fig:idp_density}, unlike the specialist force field a99SB-\textit{disp} \cite{Robustelli2018}.
The over-compaction is not as severe as it is with Amber14SB.
It is unclear whether this over-compaction arises due to training on GB3 NMR data or would arise purely when training on DFT data.
The secondary structure preference across each sequence matches experiment (Figure~\ref{fig:complex_idp}C) better than Amber14SB; this suggests that it may be suitable for the simulation of IDPs in some situations, for example when studying the binding of small molecules to IDPs \cite{Robustelli2022}.
\revision{It may also be the case that Garnet stabilises secondary structure less than Amber14SB, hence appearing better for IDPs.
Future work will use IDP data directly during training.}

\begin{figure}
  \centering
  \includegraphics[width=0.85\textwidth]{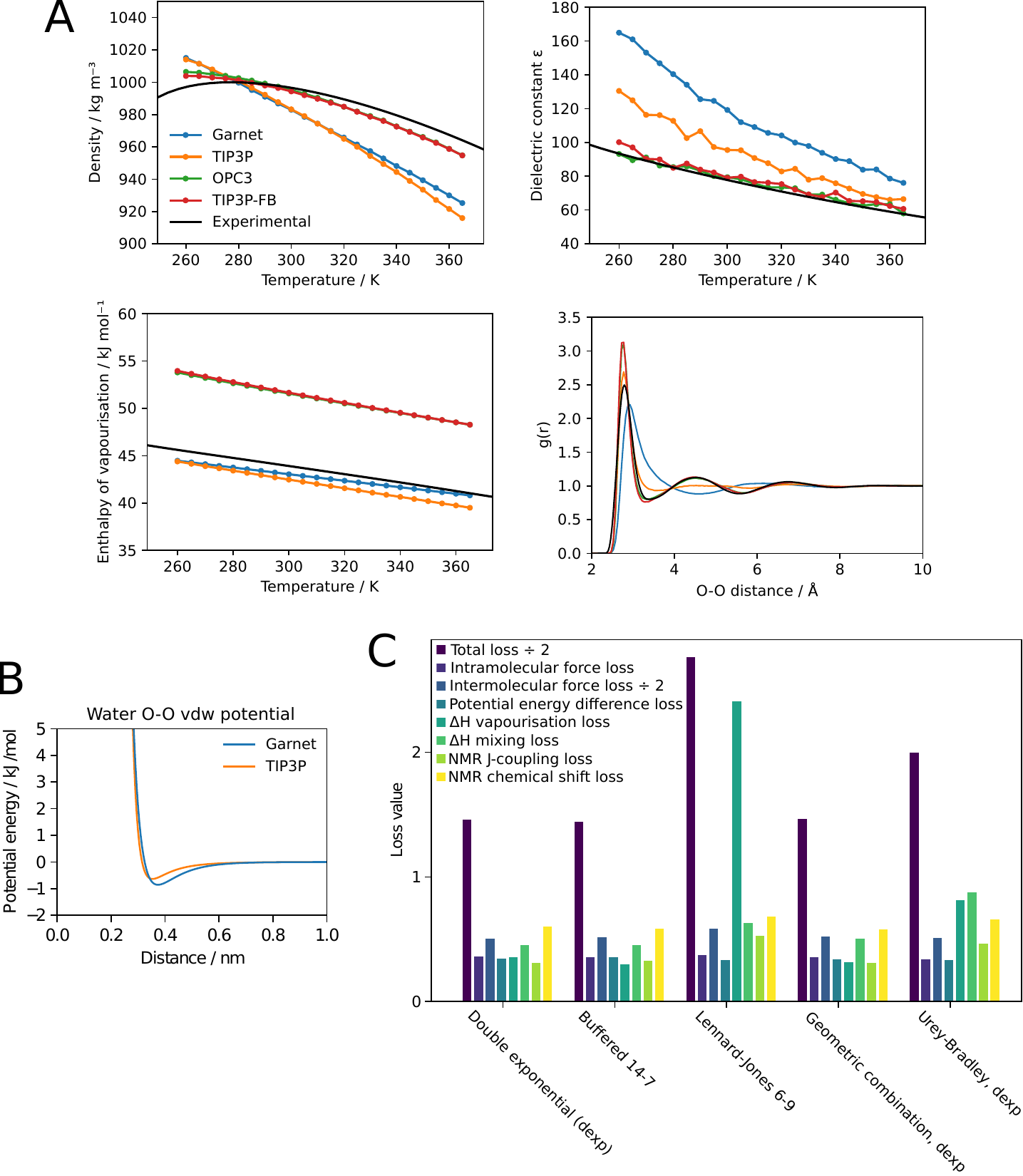}
  \caption{Garnet water model and training with different functional forms. (A) Properties of the water model learned as part of the Garnet force field, compared to TIP3P \cite{Jorgensen1983}\revision{, OPC3 \cite{Izadi2014} and TIP3P-FB \cite{Wang2014}}. Properties were calculated using the approach in OpenFF Evaluator \cite{Boothroyd2022}\revision{, so for example there is no self-polarisation correction for $\Delta H_v$}. Experimental values are from Wang et al.\ 2014 \cite{Wang2014} and Soper 2013 for the RDF \cite{Soper2013}. (B) Comparison of the water O-O double exponential interaction in Garnet and the O-O Lennard-Jones interaction from TIP3P. (C) Validation set loss values for models trained from scratch with different functional forms. For each variant, 7 models were trained and the lowest total loss of the 7 models after 11-14 epochs of training was calculated to select the best model shown here. To aid visualisation, some loss values have been scaled as noted. Double exponential corresponds to the Garnet model presented throughout the paper. Buffered 14-7 \cite{Halgren1992} and Lennard-Jones 6-9 \cite{Thurlemann2023} are alternative non-bonded potentials. Geometric combination refers to the geometric rule for combining double exponential $\sigma$ and $\varepsilon$ parameters, and Urey-Bradley refers to the Urey-Bradley bond angle potential \cite{Brooks2009}. Further functional forms were unable to train.}
  \label{fig:water}
\end{figure}

Water was not given special treatment by the model and had its parameters trained from scratch, though water was one of the molecules used for condensed phase simulations during training and is present in the DFT data.
The Garnet water parameters give bulk properties that are similar to the popular TIP3P model \cite{Jorgensen1983}, as shown in Figure~\ref{fig:water}A-B.
The dielectric constant $\varepsilon$ is too large due to the over-polarisation of charges discussed previously; MBIS gives the oxygen in the SPICE water clusters an average partial charge of -0.97, whereas in TIP3P it is -0.83.
In future, charges more relevant to the condensed phase could be used during training \cite{Adams2025}.
Alternatively, charges could not be trained directly at all or could be trained via properties like dipoles \revision{or through ensemble gradient methods.
An improved radial distribution function (RDF) could also be targeted with ensemble gradient methods to bring performance in line with modern water models like OPC3 \cite{Izadi2014} and TIP3P-FB \cite{Wang2014}.}

\subsubsection*{Relative binding free energy benchmark}

BFE methods are increasingly important in drug discovery and are now used routinely by pharmaceutical companies during lead optimisation.
A force field that treats small molecules similarly to proteins has the potential to improve the accuracy of these methods, which shows room for improvement for many targets and for freely-available force fields \cite{Baumann2026}.
We evaluated the use of Garnet for predicting binding free energies with alchemical free energy calculations. Specifically, we performed relative BFE (RBFE) calculations with Garnet parameters using the Open Free Energy (OpenFE) software, which we modified to allow simulations to be run with the double exponential potential and a related soft-core potential \cite{Horton2023}. Our evaluation was based on recent large-scale efforts to benchmark the performance of the OpenFE RBFE protocol on 58 protein-ligand systems with publicly available experimental binding affinity data \cite{Baumann2026}. The OpenFE benchmark study assessed the RBFE protocol performance using default protocol settings across all systems and used the Amber14SB \cite{Case2025} and OpenFF Sage-2.2.0 \cite{Boothroyd2023} force fields to obtain protein and small molecule parameters respectively. Briefly, OpenFE's protocol was found to rank BFEs with an accuracy comparable to that of Schrödinger’s commercial FEP+ method, although absolute deviations to experimental data were found to be smaller for FEP+ predictions. Importantly, FEP+ calculations were tuned for “maximal accuracy” results on a per-system basis \cite{ross_maximal_2023}.

Garnet was evaluated on 8 of the 58 systems from the public binding affinity benchmark set. We largely followed the default settings of OpenFE's RBFE protocol, as we aimed for our benchmark to be methodologically comparable to that of OpenFE to focus the evaluation on force field parameters rather than the BFE method. The run time for Garnet was similar to default OpenFE. We randomly picked protein-ligand systems from different ligand series categories, as previously defined by Ross et al.\ 2023 \cite{ross_maximal_2023}, to evaluate Garnet on a variety of target systems. Using OpenFE's analysis toolkit, we used the $\Delta\Delta$G$_{\mathrm{\mathrm{bind}}}$ values calculated with the RBFE protocol to estimate an absolute binding free energy ($\Delta$G$_{\mathrm{\mathrm{bind}}}$) per ligand, allowing us to (i) benchmark $\Delta\Delta$G$_{\mathrm{bind}}$ values for all possible ligand pairs of each ligand series against experimental data and (ii) evaluate if predicted $\Delta$G$_{\mathrm{bind}}$ values were ranked similarly to experiment. As in previous work \cite{Baumann2026}, we only report rank-based metrics for ligand series with more than 16 ligands and an experimental dynamic range of at least 3 kcal/mol \cite{hahn_best_2022}.

\begin{figure}
  \centering
  \includegraphics[width=1.0\textwidth]{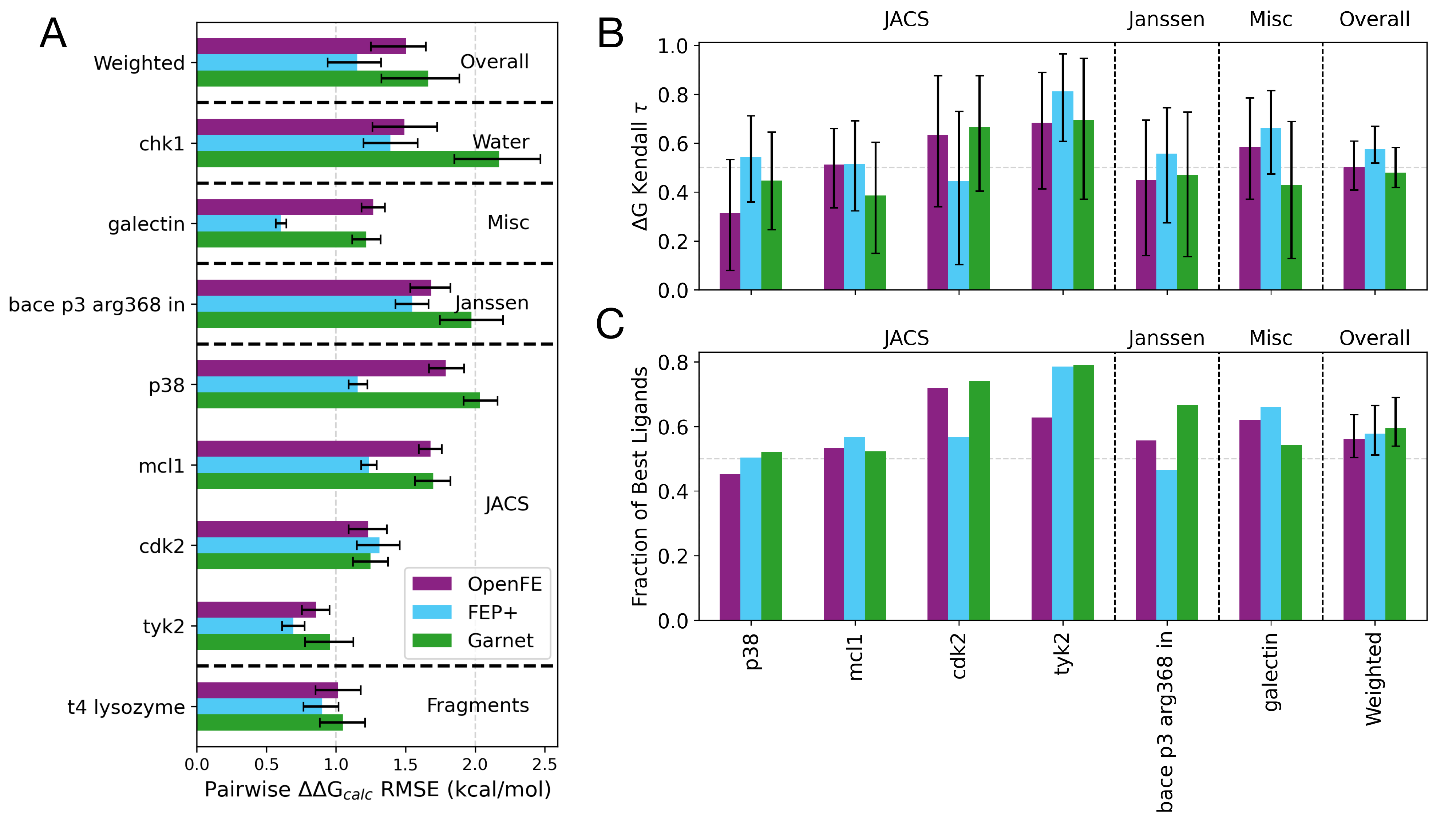}
  \caption{Performance of Garnet on relative BFE calculations. (A) RMSE between experimental and calculated $\Delta \Delta$G$_\mathrm{bind}$ values for eight proteins and associated series of ligands, with ligand series categories named as in previous work \cite{ross_maximal_2023}. $\Delta\Delta$G$_{\mathrm{bind}}$ values were calculated between all pairs of ligands in each ligand series prior to comparison to experiment. Garnet performance is compared to that of default OpenFE, which uses Amber14SB \cite{Case2025} and OpenFF Sage-2.2.0 \cite{Boothroyd2023} for protein and ligand parameters respectively, and to FEP+.  The ``Overall'' category contains the weighted RMSE across the eight benchmark systems. Error bars are bootstrapped 95\% confidence intervals. (B) Kendall's $\tau$ correlation between experimental and calculated $\Delta$G$_{\mathrm{bind}}$ values, with $\Delta$G$_{\mathrm{bind}}$ values estimated from $\Delta\Delta$G$_{\mathrm{bind}}$ as previously described \cite{Baumann2026}. A weighted average $\tau$ was calculated to summarise results. Error bars are bootstrapped 95\% confidence intervals. (C) Fraction of best ligands, evaluated by comparing $\Delta$G$_{\mathrm{bind}}$ predictions obtained from RBFE calculations to experimental $\Delta$G$_{\mathrm{bind}}$ values. FBL is reported per system and as a weighted average across systems. Kendall's $\tau$ and FBL are only reported for systems with $\geq$16 ligands and a $\Delta$G$_{\mathrm{exp}}$ dynamic range of $\geq$3 kcal/mol \cite{hahn_best_2022}. All OpenFE and FEP+ data, as well as code for plotting, were obtained from work by Baumann et al.\ 2025 \cite{Baumann2026}, with FEP+ results previously reported in Ross et al.\ 2023 \cite{ross_maximal_2023}. FEP+ calculations were tuned for “maximal accuracy” results on a per-system basis.}
  \label{fig:rbfe}
\end{figure}

We found that the RMSE between calculated and experimental $\Delta\Delta$G$_{\mathrm{bind}}$ was on average 1.66$_{1.33}^{1.89}$ kcal/mol across the eight protein targets with Garnet, compared to 1.50$_{1.25}^{1.65}$ kcal/mol for OpenFE and 1.15$_{0.94}^{1.32}$ kcal/mol for FEP+ (Figure~\ref{fig:rbfe}A, Supplementary Figure~\ref{fig:rbfe_scatter}). Garnet consistently performed comparably to OpenFE, except in the case of the chk1 outlier. Additionally, Garnet ranks ligand binding affinities relatively well, with a weighted average Kendall's $\tau$ correlation to experimental $\Delta$G$_{\mathrm{bind}}$ values similar to that of OpenFE and within error of FEP+ (Figure~\ref{fig:rbfe}B). Finally, \revision{in many cases the main aim of RBFE calculations is to prioritise ligands for further studies, and in this context it is important that the strongest binders are ranked near the top}. This can be assessed with the fraction of best ligands (FBL) metric \cite{bayly2024}. Across target systems, we found Garnet to give relatively high FBL values, and our force field appears as suited as default OpenFE and FEP+ at identifying the ligands that bind most strongly to a protein target (Figure~\ref{fig:rbfe}C). 

We noticed that ligand pairs with large deviations between $\Delta\Delta$G$_{\mathrm{bind}}$ from experiment and $\Delta\Delta$G$_{\mathrm{bind}}$ predicted by Garnet generally involved larger structural ligand changes than transformations for which the experimental $\Delta\Delta$G$_{\mathrm{bind}}$ was well-predicted. We noticed problems with net charge changes when we attempted to calculate binding free energies for series of ligands binding to the protein cmet (Merck set) and aim to solve this for the next version of our model (see Supplementary Table~\ref{tab:cmet} and Supplementary Figure~\ref{fig:cmet}). 
However, it is promising that an open, automatically parameterised force field and an open, automated BFE protocol can give competitive results to commercial alternatives.

\subsubsection*{Different functional forms}

One advantage of an automated training pipeline is that different functional forms can be assessed quickly and fairly, especially since we use our flexible Molly software for training \cite{Greener2024}.
The powerful custom interaction features in OpenMM \cite{Eastman2012} mean that a variety of functional forms are close in performance to the conventional forms, though their use in other packages may be more restricted.
As previously described, the Lennard-Jones potential gave unstable simulations during training, as did the Buckingham potential \cite{Zheng2026}.
This does not indicate that these potentials are problematic - clearly the Lennard-Jones potential is compatible with stable simulations - but does suggest that some potentials are easier to train using automated pipelines than others.
Alongside the double exponential potential, another promising non-bonded functional form was the buffered 14-7 potential \cite{Halgren1992}.
This was able to train and give stable simulations as shown in Figure~\ref{fig:water}C, with a similar loss to the double exponential potential.
The Lennard-Jones 6-9 potential \cite{Thurlemann2023} was also able to train but had a higher intermolecular force loss, suggesting that the additional global parameters in the double exponential and buffered 14-7 potentials are useful during fitting.
The geometric combination rules for double exponential parameters showed similar performance to the Lorentz-Berthelot combination rules \cite{Oliveira2023}.
The Waldman-Hagler combination rules \cite{Waldman1993} gave memory errors during training.

We tried different valence functional forms but did not find any that were better than those conventionally used.
The Morse potential gave unstable training simulations.
The Urey-Bradley angle potential \cite{Brooks2009} adds harmonic bond-like terms to the outer atoms in a bond angle and may improve the fit to vibrational spectra.
This potential was able to train but did not give distinct parameters to the harmonic bond-like term and had \revision{higher training and validation losses, suggesting that the model struggled to use the extra expressive power of the Urey-Bradley form.}

Models with different embedding dimensions were trained as shown in Supplementary Figure~\ref{fig:ablations}A.
The performance is not improved by increasing the embedding dimension from 64 to 128.
Even a model with an embedding dimension of 8 has acceptable performance on the DFT dataset.
Adding another layer to all the neural networks did not give a clear improvement.
Other GNN layers did not show improved performance over GraphSAGE \cite{Hamilton2017}; one future direction is to use the atom embeddings from \revision{existing neural networks for small molecule representation \cite{McGibbon2024}}.
We also trained models from scratch that were missing each part of the potential function in turn (Supplementary Figure~\ref{fig:ablations}B).
Removing all harmonic bonds, and to a lesser degree all harmonic angles, led to a large increase in the intramolecular force loss.
Removing torsions had little effect on the DFT losses, highlighting how training the important torsion parameters can be challenging when the loss is dominated by larger bond and angle contributions.
Explicit torsion sampling may be useful here \cite{Horton2022}.
Removing the van der Waals terms led to a large increase in the intermolecular force loss, showing that there is some signal in DFT data that can be used to train these weaker interactions.

\section*{Discussion}

One possible issue with using machine learning to find continuous atom types is the risk of overfitting, with worse performance on unseen molecules.
We deliberately made choices that would reduce this risk.
For example, the receptive field of the GNN is two, meaning that each atom embedding (continuous type) can only be influenced by atoms up to two bonds away as shown in Figure~\ref{fig:method}A.
This is conceptually similar to existing atom typing schemes, for example the SMARTS scheme used by OpenFF \cite{Wang2024}.
One consequence is that amino acids have the same Garnet parameters wherever in a protein sequence they are, provided that they are not terminal residues or followed by a proline.
This is similar to conventional residue templates.
The neural networks used are shallow with only a single hidden layer.
The atom embedding part of our model has 23,488 parameters and in total the model has 207,187 parameters.
For comparison, the CHARMM36 file \cite{Brooks2009} in OpenMM has 143,623 parameters for 635 monomers/molecules, with further molecules requiring additional parameters.
It is also worth noting that the automated training is not biased towards a particular molecule type, beyond the choice of training systems.
Performance on IDPs, for example, seems better than many force fields \cite{Robustelli2018} despite IDP data not being used during training.
This may reflect a human bias to keep proteins stably folded during manual force field tuning.
Of course, validation on other systems may reveal areas of poor performance that need targeting in future.
We noticed, for instance, that planar aromatic rings would occasionally deviate from planarity during simulations.

In this work we found that the double exponential potential trained effectively, whereas the conventional Lennard-Jones potential presented challenges.
In future, other functional forms such as class II force fields that use more complex valence terms could be trained, along with CMAP torsion corrections or more sophisticated torsion potentials.
Treatment of 1-4 interactions could be developed beyond a simple weighting on the non-bonded interactions \cite{Abdullah2025}, and different switching functions for the non-bonded interaction could be tried.
Further dispersion terms could be added \cite{Shirts2007}, or naturally soft-core non-bonded potentials could be used for greater compatibility with BFE methods.
Symbolic regression could be used to learn flexible bonded or non-bonded potentials that retain the speed of simple functional forms.
MLIPs could be used to generate data from larger systems to use during training, since Table~\ref{tab:dft_comparison} shows that they still match DFT data much better than MM force fields.
Systematic addition of features such as polarisation and charge flux \cite{Hagler2019} will allow us to assess the benefit of each addition and weigh it against the computational cost.

Force fields are often validated on BFE data, but it is rare that BFE data are used during training due to free energies being calculated from trajectories \cite{Setiadi2024, Rufa2025}.
An extension of the ensemble reweighting approach to take gradients through free energy estimators, combined with alchemical simulations in our Molly.jl software, would allow direct optimisation of force field parameters to improve BFE predictions.
More broadly, the ensemble reweighting approach could be made more accurate by using methods such as the multistate Bennett acceptance ratio \cite{Shirts2008, Rocken2024}.

The next step is to extend the systems which are well-described by the model to all species of interest to biologists: nucleic acids, lipids, metals, carbohydrates, IDPs and other polymers.
Automated training of force fields with flexible functional forms will allow us to realise the vision of universal, accurate force fields.
Since MM force fields are likely deficient in terms of both functional form and parameters, automated training provides a fast way to address both issues.
Much attention is being focused on the development of MLIPs, but MM force fields should be targeted for improvement with machine learning techniques as well.
After all, they show many of the chemical principles that MLIPs struggle to \cite{Esders2025} and are still orders of magnitude faster to run.

\section*{Methods}

\subsubsection*{Neural network architecture}

The neural network to predict MM force field parameters from molecular topologies is similar to previous methods such as Espaloma \cite{Takaba2024, Wang2022}.
Initially, a GNN converts atom input features to atom embeddings.
These embeddings are then processed by fully-connected neural networks (FCNNs) to obtain the MM force field parameters as shown in Figure~\ref{fig:method}A.

Molecules to be parameterised are processed using OpenFF Toolkit \cite{Mobley2018}.
Formal charge is assigned to individual atoms.
This presents an issue since atoms that are equivalent in the graph defined by the bonding topology, for example the two oxygen atoms in the aspartate carboxylate group, may have different formal charges.
To address this we spread the formal charge on each atom across all equivalent atoms in the topology graph.
\revision{Equivalent atoms are found by considering pairs of atoms connected by fewer than five bonds, and checking whether the two graphs remaining after each atom is removed are isomorphic.}
In the above case, each oxygen would have a spread charge of -0.5.
This ensures that equivalent atoms end up with equivalent embeddings after the GNN layers.

The atom (node) input features are the one-hot encoded element, the positive spread charge, the negative spread charge, whether the atom is aromatic, and the one-hot encoded number of bonded atoms (0 to 6).
Supported elements are those in SPICE (H, Li, B, C, N, O, F, Na, Mg, Si, P, S, Cl, K, Ca, Br, I), but performance has not been validated for all elements.
The edges are between atoms in a covalent bond and do not carry features such as bond order.
The GNN is two GraphSAGE convolution layers \cite{Hamilton2017} with the hidden layer having a width of 128 and ReLU activation.
The output atom embedding has 64 dimensions.
Since each layer can pass information through one bond, the GNN having two layers means that only atoms up to two bonds away can influence the embedding of a given atom.
The GNN does not take in conformation-dependent data like coordinates, meaning that the parameters do not change over a simulation and only need to be assigned once at the start.
This also means that stereoisomers have the same parameters.

All FCNNs used have two layers with the hidden layer having a width of 128 and ReLU activation.
Only the input and output dimensions vary between FCNN architectures.
The atom feature FCNN predicts 4 features per atom from the atom embedding, $f_1$ to $f_4$.
Partial charge is calculated following the procedure in Espaloma \cite{Wang2022}.
Two of the atom features correspond to electronegativity $e_i$ and hardness $s_i$ on atom $i$, where $E$ is the potential energy and $q_i$ is the partial charge:
$$
f_1 = e_i = \frac{\partial E}{\partial q_i}, \quad
f_2 = s_i = \frac{\partial^2 E}{\partial q_i^2}
$$
The partial charges are then calculated, where $Q$ is the total charge of the molecule containing atom $i$:
$$
q_i = -e_i s_i^{-1} + s_i^{-1} \frac{Q + \sum_i e_i s_i^{-1}}{\sum_j s_j^{-1}}
$$
The double exponential $\sigma$ parameter is calculated as sigmoid($f_3$) * 0.45 nm + 0.05 nm, meaning it can range from 0.05 nm to 0.5 nm.
The $\varepsilon$ parameter is calculated as sigmoid($f_4$) * 1.48 kJ/mol + 0.02 kJ/mol, meaning it can range from 0.02 kJ/mol to 1.5 kJ/mol.
In contrast to water models where the hydrogen atoms do not take part in Lennard-Jones interactions ($\varepsilon$ = 0 kJ/mol), here all atoms are included.
These values are wide enough to cover typical ranges for these parameters.
Analogous ranges are used for the $\alpha$ and $\beta$ global parameters and for other non-bonded functional forms.

Bond, angle, proper and improper torsion embeddings are obtained from atom embeddings in an order-independent manner using Janossy pooling \cite{Murphy2019, Wang2022}.
For bonds, the embeddings of atoms $i$ and $j$ are concatenated and passed through the bond embedding FCNN.
The bond embedding FCNN is run again, this time with the order of concatenation changed ($j$ then $i$).
The two outputs are added together to obtain the bond embedding; since addition is order-independent, the output of the bond embedding network does not change if the atom labels are switched.
A similar principle applies to the angle ($ijk$, $kji$) and proper torsion ($ijkl$, $lkji$) embedding networks.
For the improper torsion case, three inputs need to be considered since permutations of the three non-central atoms are equivalent ($ijkl$, $ikjl$, $iljk$).

Bond, angle, proper and improper torsion FCNNs convert the respective embeddings to features.
Harmonic bond and angle parameters are calculated using the approach in Espaloma to aid training \cite{Wang2022, Vanommeslaeghe2015}.
For each bond, where $f_1$ and $f_2$ are the output of the bond feature FCNN and $b_1 = 0.5$ \AA\ and $b_2 = 3$ \AA\ represent the bounding values for the bond length:
$$
k_b = f_1 + f_2, \quad
r_0 = \frac{f_1 b_1 + f_2 b_2}{f_1 + f_2}
$$
For each angle, where $f_1$ and $f_2$ are the output of the angle feature FCNN and $b_1 = 0^\circ$ and $b_2 = 180^\circ$ represent the bounding values for the bond angle:
$$
k_a = f_1 + f_2, \quad
\theta_0 = \frac{f_1 b_1 + f_2 b_2}{f_1 + f_2}
$$
For proper torsions, the outputs of the proper torsion feature FCNN $f_1$ to $f_6$ correspond to $k$ values for 6 terms with periodicity 1 to 6.
The phases are fixed to be 0 or $\pi$ for odd or even phases respectively \cite{Zheng2025}.
For improper torsions, $k$ values for 2 terms with periodicity 1 and 2 are predicted.
The 1-4 Coulomb weighting for atoms linked by 3 bonds and global parameters of the non-bonded functional form, e.g.\ $\alpha$ and $\beta$ for the double exponential potential, are also trained.
The 1-4 weighting for the double exponential and other non-bonded functional forms was set to zero since OpenMM does not support values other than 0 or 1.
Systems with multiple copies of the same molecule only need to apply parameters to each unique molecule.
This saves time when parameterising, for example, a solvated system with many water molecules.

\subsubsection*{Neural network training}

The DFT training data consisted of force, potential energy and charge data from SPICE \cite{Eastman2023, Eastman2024}, Espaloma \cite{Takaba2024}, GEMS \cite{Unke2024} and MACE-OFF \cite{Kovacs2025}.
Though these use different DFT functionals and would not be suitable for combined use when training MLIPs, differences due to the functionals are likely to be small compared to the high error inherent in MM functional forms.
In order to ensure balanced training, the datasets were used with different frequencies as shown in Supplementary Table~\ref{tab:dataset}.
Conformations from 3,000 molecules ($\sim 3$\%) were retained as a validation set to track progress during training and another 3,000 as a test set used in Table~\ref{tab:dft_comparison}.
In order to favour learning of the weak but important intermolecular forces, we split the DFT forces into intramolecular and intermolecular contributions using the approach from the MACE-OFF validation \cite{Kovacs2025, Magdau2023}.
In brief, this considers the overall translation and rotation force on the molecule as arising from intermolecular forces, and the remainder as arising from intramolecular forces.
These contributions were used to calculate separate losses, meaning that a larger loss weight could be applied to the intermolecular forces.

It is not possible to train to match DFT potential energies directly since MM force fields do not provide a meaningful absolute potential energy value.
We avoid this problem by training on the difference in potential energy between pairs of conformations of the same system.
The pairs are selected randomly each epoch.
The Adam optimiser with a learning rate of $10^{-4}$ was used.
Gradients were clamped to be between $-10^3$ and $10^3$.
For DFT training, a batch size of 256 was used and conformations containing an atom with a DFT force magnitude greater than $10^4$ kJ mol\textsuperscript{-1} nm\textsuperscript{-1} were skipped.
Expensive GEMS conformations were trained in separate batches to the other DFT data.

In addition to training on DFT data, we trained on condensed phase properties for a few systems.
Enthalpy of vapourisation data across temperatures was used for water, methanol and benzene.
Enthalpy of mixing data \cite{Boothroyd2023} was used for four varied systems: C1COCCN1/O, CN1CCCC1=O/ClCCCl, CCCCO/OC1=NCCC1 and C=CCCCC/CCCO.
As described below, each loss evaluation involved sampling a single state rather than averaging over all snapshots.
Condensed phase simulations involved simulating a 3 nm box of the relevant molecule(s) at liquid density for 2.5 ns with a 2 fs time step.
For gas phase simulations, a single molecule was simulated with no periodic boundaries.
Snapshots were taken every 10 ps and the first 1 ns was not used.
Other simulation parameters are as below for protein validation.

We also used the ensemble reweighting gradient method \cite{Wang2014, Thaler2021, Norgaard2008} to train on NMR data for the GB3 protein \cite{Cavender2025}.
This allows gradients of a loss value $L$ that depends on coordinates $x$ with respect to parameters $\lambda$ to be computed, even when the trajectory was generated using a different set of parameters \cite{Henin2022}:
$$
\frac{\mathrm{d} \left< L(x) \right>}{\mathrm{d} \lambda} = -\beta \frac{\tilde{Z}}{Z} \left( \left< L(x) D(x, \lambda) \frac{\partial U(x, \lambda)}{\partial \lambda} \right>_\sim - \allowbreak
\frac{\tilde{Z}}{Z} \left< L(x) D(x, \lambda) \right>_\sim \left< \frac{\partial U(x, \lambda)}{\partial \lambda} D(x, \lambda) \right>_\sim \right)
$$
where $D(x, \lambda) = e^{-\beta (U(x, \lambda) - \tilde{U}(x))}$, $\frac{\tilde{Z}}{Z} = 1 / \left< D(x, \lambda) \right>_\sim$, $U$ is the potential energy function, $\tilde{U}$ is the potential energy function used to generate the trajectory, $\beta = 1 / NRT$, $N$ is the number of atoms, $R$ is the molar gas constant, $T$ is the temperature and $\left< ... \right>_\sim$ represents an average over snapshots of the trajectory.
An additional $\frac{\tilde{Z}}{Z} \left< \frac{\partial L(x)}{\partial \lambda} \right>$ term is zero since the losses used do not depend directly on $\lambda$.
Due to a mistake noticed after training, the actual code used differed from the above equation as described in the source code.

Since these calculations are considerably more expensive than training on DFT data due to the larger systems and multiple snapshots, we calculate gradients using a random subset of 200 of the available snapshots every 100 batches of DFT training.
These gradients are then applied after each batch of DFT training until the next gradient update in 100 batches.
This interspersed training schedule, shown in Figure~\ref{fig:method}B, allows efficient training whilst balancing the effects of training on DFT and simulation data.
Protein simulations were run for 30 ns with a 2 fs time step.
Snapshots were taken every 10 ps and the first 5 ns was not used.
NMR data was retrieved from the Biological Magnetic Resonance Data Bank (BMRB) \cite{Hoch2023}.

At the start of training, the force field is too unstable to generate the trajectories required for the condensed phase and protein training.
Consequently, for the first 5 epochs trajectories generated using GAFF/TIP3P (for condensed phase) and Amber14SB/TIP3P (for protein) were used to provide conformations.
Technically this means that the force field has some dependence on existing force fields, but only via generated trajectories and only early in training.
From epoch 6, equivalent simulations from the training force field were used to ensure relevant conformations.
In order to not hold up training at the end of each epoch whilst the simulations are run, trajectories from two epochs ago are used.
For example, epoch 6 training used trajectories generated with the force field after epoch 4, whilst trajectories generated with the force field after epoch 5 were run at the same time on a different machine.

The losses used for the model were as follows, where $n$ is the number of atoms in a conformation:
\begin{itemize}
  \item Intramolecular force DFT loss: $\frac{0.001}{n} \sum_{i=1}^n \left\| \mathbf{F}_i - \mathbf{F}_i^{\mathrm{DFT}} \right\|$, where $\mathbf{F}_i$ is the model intramolecular force on atom $i$ and $\mathbf{F}_i^{\mathrm{DFT}}$ is the DFT intramolecular force loss on atom $i$.
  \item Intermolecular force DFT loss: $\frac{0.02}{n} \sum_{i=1}^n \left\| \mathbf{F}_i - \mathbf{F}_i^{\mathrm{DFT}} \right\|$, where $\mathbf{F}_i$ is the model intermolecular force on atom $i$ and $\mathbf{F}_i^{\mathrm{DFT}}$ is the DFT intermolecular force loss on atom $i$.
  \item Potential energy difference DFT loss: $0.01 \left | \Delta U - \Delta U^{\mathrm{DFT}} \right |$, where $\Delta U$ is the model potential energy difference between two conformations of the same system and $\Delta U^{\mathrm{DFT}}$ is the DFT potential energy difference between the same two conformations.
  \item Partial charge DFT loss: $\frac{100}{n} \sum_{i=1}^n \left | q_i - q_i^{\mathrm{DFT}} \right |$, where $q_i$ is the model partial charge on atom $i$ and $q_i^{\mathrm{DFT}}$ is the MBIS partial charge \cite{Verstraelen2016} on atom $i$ from SPICE.
  \item Enthalpy of vapourisation loss: $0.2 \left | \Delta H_v - \Delta H_v^{\mathrm{EXP}} \right |$, where $\Delta H_v$ is the model enthalpy of vapourisation and $\Delta H_v^{\mathrm{EXP}}$ is the experimental enthalpy of vapourisation. $\Delta H_v$ is calculated using the approach in OpenFF Evaluator \cite{Boothroyd2022}: $\left< U_g \right> - \left< U_l \right> + RT$, where $U_g$ is the gas phase potential energy, $U_l$ is the liquid phase potential energy per molecule, $R$ is the molar gas constant and $T$ is the temperature. However, instead of calculating the average $\left< U_l \right>$ the loss is calculated for each liquid snapshot at a time, using the average of 10 random gas phase snapshots to estimate $\left< U_g \right>$.
  \item Enthalpy of mixing loss: $0.4 \left | \Delta H_m - \Delta H_m^{\mathrm{EXP}} \right |$, where $\Delta H_m$ is the model enthalpy of mixing and $\Delta H_m^{\mathrm{EXP}}$ is the experimental enthalpy of mixing \cite{Boothroyd2023}. Three simulations were run: one of the first component, one of the second component and one of an equimolar mixture of the two components. $\Delta H_m$ is calculated using the approach in OpenFF Evaluator \cite{Boothroyd2022}:
  $$
  \Delta H_m = \frac{1}{\beta} \left( \frac{\left< u_\mathrm{mix} \right>}{N_\mathrm{mix}} - \frac{\left< u_1 \right>}{2 N_1} - \frac{\left< u_2 \right>}{2 N_2} \right)
  $$
  where $N$ is the number of molecules, $u = \beta (U + pV)$, $U$ is the potential energy, $p$ is the pressure, $V$ is the volume, $\beta = 1 / RT$, $R$ is the molar gas constant and $T$ is the temperature. However, instead of calculating each average $\left< u \right>$ the loss is calculated for snapshots from the three simulations at a time.
  \item GB3 NMR $J$-coupling ensemble reweighting loss: $0.01 \chi^2$.
  $^3J_{HN,HA}$ values were estimated using the approach in Espaloma \cite{Takaba2024, Cavender2025}:
  $$
  J(\theta) = A \cos^2(\theta + \Delta) + B \cos(\theta + \Delta) + C
  $$
  where $\theta$ is the associated torsion angle and $A = 7.97$, $B = -1.26$, $C = 0.63$ and $\Delta = 60^\circ$ are empirical Karplus parameters.
  Agreement with experiment was assessed by computing $\chi^2$ values:
  $$
  \chi^2 = \frac{1}{N} \sum_{i=1}^N \frac{(J - J^{\mathrm{EXP}})^2}{\sigma^2}
  $$
  where the sum is over $N$ observables, $J^{\mathrm{EXP}}$ is the experimental $J$-coupling \cite{Hoch2023} and $\sigma = 0.42$ is the systematic error in the Karplus model.
  \item GB3 NMR chemical shift ensemble reweighting loss: $\frac{0.5}{n} \sum_{i=1}^{n_b} \left | \delta_i - \delta_i^{\mathrm{EXP}} \right |$, where $n_b$ is the number of backbone CA/C/N/H atoms, $\delta_i$ is the model chemical shift for backbone atom $i$ predicted using Graph NMR \cite{Yang2021} and $\delta_i^{\mathrm{EXP}}$ is the experimental chemical shift for backbone atom $i$ \cite{Hoch2023}.
  \item Proper torsion regularisation: $\frac{0.01}{n_t} \sum_{i=1}^{n_t} \left | k_i \right |$, where $n_t$ is the number of proper torsion $k$ values and $k_i$ is the value of proper torsion $k$ value $i$.
  \item Neural network weight regularisation: $10^{-5} \sum_{i=1}^{n_p} p_i^2$, where $n_p$ is the number of neural network parameters in the model and $p_i$ is the value of paramter $i$.
\end{itemize}

Julia was used to train the model due to its good performance and automatic differentiation (AD) support \cite{Bezanson2017, Roesch2023}.
The Molly.jl package provided many of the molecular dynamics components \cite{Greener2024}.
Enzyme.jl \cite{Moses2020} and Zygote.jl \cite{Innes2018} were used to calculate gradients with AD, including through complex algorithms such as particle mesh Ewald (PME) \cite{Darden1993}.
MDAnalysis \cite{Gowers2016} and BioStructures \cite{Greener2020} were used for analysis.
The model was trained multi-threaded on CPU with training having high memory requirements (up to 754 GB).
Single precision floats were used.
The simulations run during training were run on GPU using OpenMM \cite{Eastman2017}.

7 training runs were carried out and the best model was selected manually.
Training took around 5 days, with the model after 12 epochs treated as the final model.
A plot of the various losses during training is shown in Supplementary Figure~\ref{fig:training}.
The model was converted to PyTorch \cite{Paszke2019} to allow convenient inference by users using the trained parameters in the OpenFF \cite{Wang2024} and OpenMM \cite{Eastman2017} software ecosystems.
The output of the Julia and PyTorch networks is the same.
We have attempted to make the process of parameterising systems as simple as possible.
In general, as long as the system can be read into an OpenFF Topology object \cite{Mobley2018} and consists of compatible elements then Garnet can assign it parameters in a few seconds.

\subsubsection*{Potential functional forms}

For the simulation of condensed phase systems, the cutoff distance for non-bonded interactions was 1 nm and electrostatics were implemented using PME \cite{Darden1993}.
When training on gas phase DFT data, no cutoff was used for non-bonded interactions and the Coulomb potential was used for electrostatics.

In all cases, $r$ is the interatomic distance and $\theta$ is the bond angle.
The harmonic bond potential has parameters $k_b$ and $r_0$:
$$
V(r, k_b, r_0) = \frac{k_b}{2} (r - r_0)^2
$$
The harmonic angle potential has parameters $k_a$ and $\theta_0$:
$$
V(\theta, k_a, \theta_0) = \frac{k_a}{2} (\theta - \theta_0)^2
$$
The cosine torsion potential for proper and improper torsions has parameters $k_n$ and $\phi_{s,n}$, though the phases $\phi_{s,n}$ were not trained:
$$
V(\phi, k_n, \phi_{s,n}) = \sum_{n=1}^N k_n (1 + \cos(n \phi - \phi_{s,n}))
$$
The double exponential potential has parameters $\sigma$, $\varepsilon$, $\alpha$ and $\beta$, where $r_m = 2^\frac{1}{6} \sigma$:
$$
V(r, \sigma, \varepsilon, \alpha, \beta) = \varepsilon \left[ \frac{\beta e^\alpha}{\alpha - \beta} \exp \left( -\alpha \frac{r}{r_m} \right) - \frac{\alpha e^\beta}{\alpha - \beta} \exp \left( -\beta \frac{r}{r_m} \right) \right]
$$
The Lorentz-Berthelot combination rules were used:
$$
\sigma_{ij} = \frac{\sigma_i + \sigma_j}{2}, \quad
\varepsilon_{ij} = \sqrt{\varepsilon_i \varepsilon_j}
$$
The Lennard-Jones potential that failed to train has parameters $\sigma$ and $\varepsilon$:
$$
V(r, \sigma, \varepsilon) = 4\varepsilon \left[\left(\frac{\sigma}{r}\right)^{12} - \left(\frac{\sigma}{r}\right)^{6}\right]
$$
The results of training with different functional forms are shown in Figure~\ref{fig:water}C.
The Lennard-Jones 6-9 potential \cite{Thurlemann2023} has parameters $\sigma$ and $\varepsilon$:
$$
V(r, \sigma, \varepsilon) = \frac{27\varepsilon}{4} \left[\left(\frac{\sigma}{r}\right)^{9} - \left(\frac{\sigma}{r}\right)^{6}\right]
$$
The buffered 14-7 \cite{Halgren1992} potential has parameters $\sigma$, $\varepsilon$, $\delta$ and $\gamma$, where $\rho = r / r_m$ and $r_m = 2^\frac{1}{6} \sigma$:
$$
V(r, \sigma, \varepsilon, \delta, \gamma) = \varepsilon \left(\frac{1 + \delta}{\rho + \delta}\right)^7 \left(\frac{1 + \gamma}{\rho^7 + \gamma} - 2\right)
$$
The geometric combination rules are an alternative to the Lorentz-Berthelot combination rules:
$$
\sigma_{ij} = \sqrt{\sigma_i \sigma_j}, \quad
\varepsilon_{ij} = \sqrt{\varepsilon_i \varepsilon_j}
$$
The Urey-Bradley potential \cite{Brooks2009} replaces the harmonic angle potential and has parameters $k_a$, $\theta_0$, $k_u$ and $r_0$, where $r$ is the distance between the outer atoms:
$$
V(\theta, r, k_a, \theta_0, k_u, r_0) = \frac{k_a}{2} (\theta - \theta_0)^2 + \frac{k_u}{2} (r - r_0)^2
$$
In each case the force was implemented directly and AD was used to obtain the gradients required for training.

\subsubsection*{Small molecule benchmark}

We benchmarked Garnet against the OpenFF Industry Benchmark dataset \cite{Horton2025}. The dataset contains 9,835 unique molecules with 74,585 conformers and aims to represent a collection of chemical species that are of particular interest to industry partners.
We focused our comparison on calculating several deviation metrics from the QM-minimised structures found in the dataset, which we take as the ground truth. For each molecule, we iterated over all available QM-minimised conformers. Each QM structure was used as the starting configuration for an energy minimisation performed with the force field under consideration, using a steepest descent algorithm. This procedure isolates the structural preferences encoded by the force field while avoiding biases associated with initial conformational differences.

Following minimisation, the resulting force field-optimised structure was rigidly aligned to the corresponding QM reference conformer using the Kabsch algorithm. Deviation metrics were then computed between the aligned structures. We calculated the RMSD of several properties. Without lack of generality, the RMSD is defined as:
$$
\mathrm{RMSD}[X] = \sqrt{\frac{1}{N} \sum_{i=1}^{N} \left( X_{i}^\mathrm{FF} - X_{i}^\mathrm{QM} \right)^2}
$$
where $X_i^\mathrm{FF}$ and $X_i^\mathrm{QM}$ are the force field minimised and QM structure values for instance $i$ of a general observable $X$, and $N$ is the total number of instances. We calculated the RMSD for the full Cartesian coordinates of the molecules, as well as for bonds, angles, proper torsions, and improper torsions. In addition, we included the TFD \cite{Schulz2012}, as implemented in RDKit \cite{Landrum2025}.

We also computed an energy deviation metric to assess the relative energetic ranking of conformers. Taking advantage of the fact that each molecule in the QM dataset is represented by several conformers, we compared conformational energies relative to the global QM energy minimum of each set. Specifically, for a given molecule, we identified the conformer with the lowest QM energy and used it as a reference for both the QM and force field-minimised ensembles. Furthermore, to ensure the energetic comparison remained physically meaningful, only conformers $i$ with a structure RMSD $\le 1.0$~\AA\ relative to the QM reference were included in the evaluation. The energy deviation for a conformer $i$ is defined as:
$$
\Delta\Delta\mathrm{E}_{i} = \left(\mathrm{E}_{i}^\mathrm{FF} - \mathrm{E}_\mathrm{min}^\mathrm{FF}\right) - \left(\mathrm{E}_{i}^\mathrm{QM} - \mathrm{E}_\mathrm{min}^\mathrm{QM}\right)
$$
where $\mathrm{E}^\mathrm{FF}$ and $\mathrm{E}^\mathrm{QM}$ represent the energies of the force field minimised and QM reference structures, respectively. For each molecule, we report the average $\Delta\Delta\mathrm{E}$ across all qualifying conformers ($i \neq \mathrm{min}$). Working with energy differences relative to the QM global minimum removes the arbitrary energy offset associated with absolute energy scales, ensuring that the metric specifically assesses the force field's ability to reproduce the relative energetic ordering required for accurate thermodynamic and kinetic modelling.

A statistical analysis was performed to assess whether the quality of the force field description depends systematically on the presence of specific chemical functional groups. We annotated each molecule with a curated set of chemical motifs defined through SMARTS patterns and identified using substructure searches in RDKit \cite{Landrum2025}. A molecule was considered to contain a given motif if at least one substructure match was found.
Using the calculated deviation metrics, molecules were partitioned into subsets to isolate extreme performance regimes. For each metric, we identified the first and ninety-ninth percentiles of the error distribution to define the best-performing (lowest 1\% error) and worst-performing (highest 1\% error) tiers respectively. This procedure allows us to contrast chemical motifs preferentially associated with high-fidelity structural descriptions against those that consistently drive significant geometric or conformational deviations.

For a given motif $m$ and a target performance tier, we compared the motif's frequency within the tier to its frequency in the rest of the dataset. Let $N_{\mathrm{tier}}$ and $N_{\mathrm{rest}}$ be the total number of molecules in the tier and the complementary set, and let $a_m$ and $c_m$ be the number of molecules in each set containing the motif. The empirical motif frequencies are given by
$$
f_{\mathrm{tier}}^{(m)} = \frac{a_m}{N_{\mathrm{tier}}}, \quad
f_{\mathrm{rest}}^{(m)} = \frac{c_m}{N_{\mathrm{rest}}}
$$
From these estimates, we defined an enrichment ratio $E^{(m)} = f_{\mathrm{tier}}^{(m)} / f_{\mathrm{rest}}^{(m)}$. For the worst-performing tier, we specifically report the enrichment of errors as $E_{\mathrm{high/low}}^{(m)}$, where a ratio greater than unity indicates that a functional group is overrepresented among poorly described molecules.

To fully propagate statistical uncertainty, we adopted a Bayesian treatment of motif enrichment. Observed motif counts were interpreted as realizations of binomial random variables with success probabilities $p_{\mathrm{tier}}^{(m)}$ and $p_{\mathrm{rest}}^{(m)}$. Assuming non-informative $\text{Beta}(1,1)$ priors, the conjugate posteriors are
$$
p_{\mathrm{tier}}^{(m)} \sim \mathrm{Beta}(a_m + 1,\, N_{\mathrm{tier}} - a_m + 1), \quad
p_{\mathrm{rest}}^{(m)} \sim \mathrm{Beta}(c_m + 1,\, N_{\mathrm{rest}} - c_m + 1)
$$
We generated 20,000 samples from these distributions to compute a posterior over the enrichment ratio, from which we report the median, a 95\% credible interval, and the posterior probability that the ratio exceeds unity. This analysis was carried out for OpenFF, Espaloma, MACE-OFF, and Garnet to compare their relative performances. Results were filtered to include only significant motifs meeting the following criteria: a minimum occurrence of 10 times in the dataset, an enrichment ratio of at least 1.25, and a Bayesian probability of enrichment exceeding 0.99. This procedure was applied to the full suite of deviation metrics: structure, bond, angle, proper torsion and improper torsion RMSDs, as well as TFDs.

\revision{
Since Garnet was trained on small QM systems and MBIS charges \cite{Verstraelen2016, Eastman2023, Eastman2024}, which are known to lead to over-polarisation issues, we tested the effect of this choice by calculating the dipole moments predicted by OpenFF, Espaloma, and Garnet, and comparing them to the QM ground truth.

The data used for this was the $\sim 3$\% held-out test split of the SPICE dataset. The dipole moment is defined as:
$$
    \boldsymbol{\mu} = \sum_{i}q_i \cdot \mathbf{r}_i
$$
where $q_i$ is the partial charge and $\mathbf{r}_i$ is the position vector of atom $i$. However, the absolute orientation of each molecule and conformer in 3D space is arbitrary. Therefore, instead of comparing raw dipoles, it is more informative to compare the projection of said dipoles on the internal axes of the molecules. If we centre the coordinates at the centre of mass of each molecule, and calculate the inertial tensor $\mathbf{I}$:
$$
    \mathbf{r}^{\prime}_{i} = \mathbf{r}_i - \frac{\sum_{j}m_{j} \cdot \mathbf{r}_j}{\sum_j m_j}
$$
$$
    \mathbf{I} = \sum_{i} m_{i} \left( ||\mathbf{r}^{\prime}_i||^2\cdot\mathbf{1} - \mathbf{r}^{\prime}_i \otimes \mathbf{r}^{\prime}_i \right)
$$
where $m_i$ is the mass of atom $i$, we can obtain the internal frame of the molecule by eigendecomposition of the tensor $\mathbf{I}=\mathbf{A}\mathbf{\Lambda}\mathbf{A}^{T}$, where the columns of $\mathbf{A}$ are the principal axes. Since the sign of the eigenvectors is arbitrary, we ensure a right-handed triad by enforcing $\mathrm{det}(\mathbf{A})>0$, flipping the sign of the third axis when necessary.

We can thus project the dipole moment onto the molecular frame of reference by:
$$
    \boldsymbol{\mu}^{\prime} = \sum_{i}q_i \cdot \mathbf{r}^{\prime}_i
$$
$$
    \boldsymbol{\mu}_{body} = \mathbf{A}^{T} \boldsymbol{\mu}^{\prime}
$$
The magnitude difference, $|| \boldsymbol{\mu}_{body}^{FF}|| - || \boldsymbol{\mu}_{body}^{QM} ||$; vector difference, $|| \boldsymbol{\mu}_{body}^{FF} - \boldsymbol{\mu}_{body}^{QM} ||$; and angle between the dipoles, $\angle \left( \boldsymbol{\mu}_{body}^{FF}, \boldsymbol{\mu}_{body}^{QM} \right)$, are reported. We supplement this with the dipole tensor $\mathbf{D}=\boldsymbol{\mu}_{body}\otimes\boldsymbol{\mu}_{body}$, which is invariant under transformations $\boldsymbol{\mu}\rightarrow-\boldsymbol{\mu}$, by comparing the absolute $|| \mathbf{D}^{FF} - \mathbf{D}^{QM}  ||_{\mathrm{Frob}}$, QM-relative $\frac{|| \mathbf{D}^{FF} - \mathbf{D}^{QM}  ||_{\mathrm{Frob}}}{||\mathbf{D}^{QM}||_{\mathrm{Frob}}}$, and symmetric $\frac{|| \mathbf{D}^{FF} - \mathbf{D}^{QM}  ||_{\mathrm{Frob}}}{ \max \left[ ||\mathbf{D}^{QM}||_{\mathrm{Frob}}, ||\mathbf{D}^{FF}||_{\mathrm{Frob}} \right] }$ Frobenius norm differences between the QM and force field dipole tensors.
}

\revision{

We evaluated Garnet against $\rho$ and $\Delta H_v$ for 12 small molecules using previously curated experimental data \cite{wang_application_2011} as well as data from the NIST Chemistry WebBook \cite{NISTWebBook}. The 12 small molecules included in our benchmark set are butane, cyclohexane, 2-propanol, phenol, propane-1-thiol, acetic acid, acetone, N-methylpropanamide, trichloromethane, 4-methyl thiazole, ﬂuorobenzene, and trimethylphosphate.  Additionally, we calculated $\rho$ and $\Delta H_v$ for the three training set molecules water, methanol, and benzene to assess potential overtraining of Garnet. We used experimental $\rho$ and $\Delta H_v$ data recorded at the same temperature for individual molecules, except for phenol for which data were measured at two different temperatures. For evaluations of OpenFF-2.1.1 and GAFF, TIP3P was used to parameterise water. 

As described above, we calculated $\Delta H_v$ from simulations using the approach of OpenFF Evaluator \cite{Boothroyd2022}, meaning that we separately simulated a liquid phase and a gas phase of each of the 15 molecules in our small molecule set (12 test molecules and three training molecules). Liquid phase simulations were initialised in a 3 nm cubic box with a mass density equal to the experimental mass density, and simulations were run with a Monte Carlo barostat to maintain a pressure of 1 bar. Single molecule systems with no periodic boundary were used to simulate the gas phase. All systems were simulated with Langevin dynamics to maintain a temperature equal to the temperature at which experimental $\rho$ and $\Delta H_v$ data were collected. For phenol, experimental $\rho$ and $\Delta H_v$ data were not reported for the same temperature, and we therefore simulated phenol at the two different temperatures. Bonds to hydrogen were constrained. In liquid phase simulations, a distance cutoff of 1 nm was applied to non-bonded interactions, and electrostatic interactions were evaluated with PME. All simulations were run for 200 ns using a time step of 2 fs and saving simulation snapshots every 10 ps. The first 5 ns of all trajectories was discarded prior to analysis.

$\Delta H_v$ was calculated using ensemble-averaged potential energies from liquid and gas phase simulations according to the equation $\left< U_g \right> - \left< U_l \right> + RT$, where $U_g$ is the gas phase potential energy, $U_l$ is the liquid phase potential energy per molecule, $R$ is the molar gas constant and $T$ is the temperature \cite{Boothroyd2022}. $\rho$ was calculated from the liquid phase simulations. We noticed that GAFF simulations of cyclohexane and acetic acid did not converge with respect to $\rho$ even with extended simulations of 400 ns, and we hence did not include simulation data for these two molecules in our evaluation of GAFF. 

}

\subsubsection*{Protein benchmark}

For the four folded proteins and four protein complexes, three independent replicas of each system were simulated for 5~\textmu s each. All systems were kept at a constant temperature of 300 K using a Langevin integrator with a friction coefficient of 1 ps\textsuperscript{-1}, and coupled to a Monte Carlo barostat in order to maintain a constant pressure of 1 bar. The cutoff distance for non-bonded interactions was 1 nm and electrostatics were implemented using PME \cite{Darden1993}. The internal degrees of freedom for water molecules were removed, and constraints were applied to bonds involving hydrogen atoms. This, in addition to using hydrogen mass repartitioning (HMR) \cite{Hopkins2015} to increase the hydrogen mass to 2 atomic units, allowed the simulations to run with a time step of 4 fs. We noticed that Garnet simulations were stable with a 4 fs time step without HMR, whereas Amber14SB simulations were not.

These proteins have been extensively characterised by NMR spectroscopy, and we focused on scalar coupling constants ($J$-couplings), which can be estimated from simulation trajectories through the Karplus relationships. We computed several $^3J$-couplings corresponding to atoms connected through three consecutive covalent bonds, as well as scalar couplings arising from through-space interactions mediated by backbone hydrogen bonds. The calculated couplings were compared against a curated set of experimental measurements \cite{Cavender2022}. For more details, see Takaba et al.\ 2024 \cite{Takaba2024}.

All NMR observables were computed by post-processing trajectory snapshots taken at 500 ps intervals after discarding an initial equilibration period of 1~\textmu s. At each saved frame, coordinates were extracted, converted into scalar coupling estimates, and accumulated over time to obtain an ensemble of predicted couplings. For covalently mediated couplings, we evaluated $^3J$-couplings using standard protein torsion definitions ($\phi$ and $\chi_1$) and a Karplus functional form:
$$
^3J(\theta) = A\cos^2(\theta + \delta) + B\cos(\theta + \delta) + C,
$$
with parameters $\delta$, $A$, $B$ and $C$ selected according to the atom quartet and residue identity. In addition, we computed the minimum and maximum values attainable by the corresponding Karplus curve and clamped the experimentally reported coupling to this theoretically admissible interval prior to analysis, \revision{since many values lie outside the range of the relevant curves \cite{Maier2015}}.

For hydrogen bond-mediated couplings (GB3 and Ubq), we employed a curated list of donor--acceptor residue pairs. For each pair and frame, a backbone N--H$\cdots$O=C hydrogen bond was identified using a geometric criterion (distance cutoff and maximum donor--hydrogen--acceptor angle). When present, the hydrogen bond geometry was used in the corresponding Karplus-type expression:
$$
J_{\mathrm{HB}}(R,\theta,\phi) = \exp\!\left[-k\,(R - R_0)\right] \left[ \left( A \cos^2 \phi + B \cos \phi + C \right)\sin^2 \theta + D \cos^2 \theta \right],
$$
where $R$ is the H$\cdots$O distance, $\theta$ is the H--O--C angle, and $\phi$ is the H--O--C--N torsion angle. The acceptor nitrogen was defined as the amide nitrogen covalently bound to the acceptor carbonyl group.

Following individual calculations, ensembles were post-processed using a moving block bootstrap procedure with 2,000 samples to estimate uncertainty. We evaluated the following estimators:
\begin{itemize}
    \item Root mean square error (RMSE): used for hydrogen-bond mediated $^3J_{\mathrm{N,C'}}$-couplings to quantify absolute deviation in Hz.
    \item Absolute normalised error (ANE) \revision{\cite{Takaba2024, Maier2015}}: used for torsion-mediated couplings to compare different $^3J$ types independently of their specific magnitude:
    \[
    \mathrm{ANE} = \frac{|J_{\mathrm{calc}} - J_{\mathrm{exp,clamped}}|}{J_{\mathrm{max}} - J_{\mathrm{min}}}
    \]
\end{itemize}
Mean values and 99.9\% confidence intervals were determined and reported across all proteins and models. In the case of HEWL, backbone couplings were omitted due to the absence of corresponding experimental measurements.

\subsubsection*{Relative binding free energy benchmark}

We performed alchemical free energy calculations using OpenFE \cite{Baumann2026}. Briefly, to estimate RBFEs between two ligands, the OpenFE RBFE protocol creates a hybrid topology ligand and simulates the hybrid ligand in solvent or bound to a receptor protein at a number of $\lambda$ states ($0 \le \lambda \le 1$), where $\lambda$ determines the degree to which alchemical (i.e.\ mutated) atoms of the hybrid ligand interact with other ligand atoms and the rest of the system.

The RBFE protocol of OpenFE uses the Lennard-Jones potential and is not immediately compatible with other potentials such as the double exponential potential. We modified the RBFE protocol source code to be able to perform simulations with the double exponential potential and a related soft-core potential \cite{Horton2023}. Our modified version of the RBFE protocol is available and is compatible with any custom potential for which $\sigma$ and $\varepsilon$ are the only per-atom force field parameters, in addition to charge.

For simulations at intermediate alchemical states ($0 < \lambda < 1$), we used a soft-core double exponential potential to describe interactions involving alchemical atoms:
$$
V(r, \sigma, \varepsilon, \alpha, \beta, \lambda) = \lambda \varepsilon
\left[ \frac{\beta_{s} e^ {\alpha_{s}}}{\alpha_{s} - \beta_{s}} \exp \left( -\alpha_{s} \frac{r}{r_m} \right) - \frac{\alpha_{s} e^{\beta_{s}}}{\alpha_{s} - \beta_{s}}
\exp \left( -\beta_{s} \frac{r}{r_m} \right) \right]
$$
where $r_m = 2^\frac{1}{6} \sigma$, $\alpha_{s} = (1.1 + \lambda(\alpha - 1.1))$ and $\beta_{s} = (1 + \lambda(\beta - 1))$ \cite{Horton2023}. $\alpha$, $\beta$, $\sigma$ and $\varepsilon$ are the same parameters as in the standard double exponential potential and were predicted with Garnet. At $\lambda=1$, the potential reduces to the regular double exponential potential. Electrostatic interactions were not treated with a soft-core potential. 

We ran OpenFE's RBFE protocol using default openfe-v1.8.0 simulation and alchemical transformation settings, with the exception that we linearly interpolated electrostatic interactions involving alchemical atoms separated by three bonds (1-4 interactions) between $\lambda$ states. In this way, 1-4 interactions were switched off and on at $\lambda = 0$ and $\lambda = 1$ respectively and linearly interpolated between, similar to all other electrostatic interactions.

Following openfe-v1.8.0 default settings, all simulations were run at 298.15 K and 1 bar in a dodecahedral simulation box with 1.5 nm of solvent padding, using Na$^+$ and Cl$^-$ as neutralising ions and to give an ion concentration of 150 mM. 11 $\lambda$ windows were used per ligand transformation, except when a transformation involved ligand net charge changes, in which case 22 $\lambda$ windows were used. Following 5,000 steps of minimization, 1 ns of equilibration and 5 ns (or 20 ns in the case of net charge changes) of production simulation were run per $\lambda$ window. \revision{During net charge-changing ligand transformations, the system net charge was kept constant by introducing an alchemical water molecule that was gradually transformed into a counter ion to oppose the charge change caused by the ligand transformation.} All transformations were run in replicates of three and using Hamiltonian replicate exchange to enhance sampling.

We used pre-prepared protein and ligand input structures from the OpenFE Industry Benchmarking Project (v1.0.0) \cite{Baumann2026, Baumann2025b}. These structures were originally obtained by OpenFE from the Schr\"{o}dinger 2023 benchmark set (v2.0) \cite{ross_maximal_2023} and modified to be OpenFE-compatible \cite{Baumann2026}. As in previous work \cite{Baumann2026}, alchemical transformation networks were planned with the Lead Optimization Mapper algorithm \cite{liu_lead_2013}, and atom mappings were calculated with Kartograf \cite{ries_kartograf_2024}. All protein, ligand and solvent molecules were parameterised with Garnet. For further details on the RBFE protocol and simulation setup, we refer to the OpenFE Industry Benchmarking Project paper and GitHub \cite{Baumann2026,Baumann2025b}.

Inspired by Takaba et al. 2024 \cite{Takaba2024}, we first evaluated Garnet for RBFE calculations using protein-ligand data for Tyk2, Cdk2, P38 and Mcl1. Subsequently, non-JACS systems were selected randomly from the different ligand series categories used by Ross et al.\ 2023 \cite{ross_maximal_2023}, though using the criteria that we did not simulate the same protein twice even if it occurred in two separate ligand series categories, and we additionally avoided systems with few ligands ($<$10) and many ligands ($>$40). 

We benchmarked RBFEs calculated with Garnet against experimental data that was made available recently via the OpenFE Industry Benchmarking Project GitHub \cite{Baumann2026,Baumann2025b} and originally by Ross et al.\ 2023 \cite{ross_maximal_2023}. Using the OpenFE \textit{gather} command line option, we performed maximum likelihood estimation to calculate a $\Delta G_{\mathrm{bind}}$ per ligand in each protein system \cite{xu_optimal_2019}, using as input for the estimator the $\Delta \Delta G_{\mathrm{bind}}$ values calculated with the RBFE protocol for the ligand transformations present in the alchemical transformation network. We used the $\Delta G_{\mathrm{bind}}$ scores to calculate a $\Delta \Delta G_{\mathrm{bind}}$ for all possible ligand pairs in a ligand series (that is, per protein) and used these $\Delta \Delta G_{\mathrm{bind}}$ values to estimate RMSEs to experimental results. We used $\Delta G_{\mathrm{bind}}$ values to calculate correlation metrics to the experimental data, including Kendall's $\tau$ correlation between calculated and experimental $\Delta G_{\mathrm{bind}}$ and the FBL. Generally, to estimate the FBL for a given ligand series, the fractional overlap between the $x$ ``best ligands'' (i.e.\ the $x$ strongest binders) from experiment and from predictions is calculated for all values of $x$ up to 50\% of the total number of ligands in the series, and the average fractional overlap is then reported \cite{bayly2024}. 

To evaluate the performance of Garnet across the eight protein systems, we calculated a weighted RMSE:
$$
\mathrm{RMSE} = \sqrt{\frac{1}{\sum_{i}^{M} n_{i}} \sum_{i}^{M} n_{i} \mathrm{RMSE}_{i}^2},
$$
where $M$ is the number of systems and system $i$ has $n_i$ ligands with $\mathrm{RMSE}_{i}$ to the experimental data \cite{Baumann2026}. We similarly calculated weighted averages of Kendall's $\tau$ correlations and FBL values across $M$ systems according to
$$
\tau = \frac{1}{\sum_{i}^{M} n_{i}} \sum \tau_{i} \cdot n_{i},
$$
and 
$$
\mathrm{FBL} = \frac{1}{\sum_{i}^{M} n_{i}} \sum \mathrm{FBL}_{i} \cdot n_{i}
$$

We generally quantified uncertainties in our benchmarking using 95\% confidence intervals, which we estimated with bootstrapping. For a given evaluation metric (RMSE, Kendall's $\tau$ or FBL), we resampled the relevant input data for the performance evaluation with replacement 1,000 times to estimate a distribution of performance scores and reported the 2.5th and 97.5th percentiles of the distribution as the 95\% confidence interval. 

We compared predictions of RBFEs with Garnet to predictions obtained with the default OpenFE RBFE protocol. These predictions were calculated as part of the OpenFE Industry Benchmarking Project and reported by Baumann et al. 2025 \cite{Baumann2026}. The OpenFE protocol parameterises proteins with Amber14SB \cite{Case2025}, water molecules with TIP3P \cite{Jorgensen1983}, and small molecules (and cofactors) with Open Force Field Sage-2.2.0 \cite{Boothroyd2023} and AM1-BCC \cite{jakalian2000}. We note that we have used openfe-v1.8.0 and that the OpenFE Industry Benchmarking Project was run with openfe-v1.0.1. We also compared our results to $\Delta \Delta G_{\mathrm{bind}}$ predictions from FEP$+$ \cite{ross_maximal_2023}.

Some ligand transformations failed consistently across simulation replicates when we ran calculations for bace p3 arg368 in and cmet. For bace p3 arg368 in, failed transformations all involved an end-state ligand containing a single carbon-carbon triple bond that needed to be created during the alchemical transformation. For cmet, failed transformations involved net charge changes. Since the failed edges were redundant in the ligand transformation network, i.e.\ there was a path in the network between all ligands even if these edges were excluded, we removed these edges from the transformation network for data analysis.

\begin{table}
    \centering
    \begin{tabular}{c c c}
        \hline
        \textbf{Dataset name} & \textbf{Protein} & \textbf{n ligands} \\
        \hline
        JACS & tyk2 & 16 \\
        JACS & cdk2 & 16 \\
        JACS & p38 & 34 \\
        JACS & mcl1 & 42 \\
        Water & chk1 & 13 \\
        Misc & galectin & 26 \\
        Janssen & bace p3 arg368 in & 21 \\
        Fragments & t4 lysozyme & 12 \\
        Merck & cmet & 24 \\
    \end{tabular}
    \caption{Protein-ligand systems used for RBFE benchmarking of Garnet. The dataset name follows naming by Ross et al.\ 2023 \cite{ross_maximal_2023}. The number of ligands in the ligand series is indicated for each protein target.}
    \label{tab:placeholder}
\end{table}

\subsection*{Availability}

The Garnet force field, training scripts, validation scripts and data are available under a permissive licence at \url{https://github.com/greener-group/garnet}. The modified RBFE protocol is available at \url{https://github.com/greener-group/openfe}.
Molly.jl is available at \url{https://github.com/JuliaMolSim/Molly.jl}.

\subsection*{Author contributions}

\revision{JGG conceived the idea and trained the model.
ABG, TKS and JGG ran the benchmarks and wrote the paper.
ER assisted with the free energy benchmark.
All authors discussed the results and contributed to the final manuscript.}

\subsection*{Conflict of interest}

The authors declare no competing interests.

\subsection*{Acknowledgements}

We thank Giuseppe Gambini, Richard Geng, Daniel Cole, Joshua Horton, Finlay Clark, Chapin Cavender, Yuanqing Wang, the Sjors Scheres group and the OpenFF community for useful discussions; William Moses and Valentin Churavy for support with Enzyme.jl; and Jake Grimmett, Toby Darling and Ivan Clayson for help with high-performance computing.
This work was supported by the Medical Research Council, as part of United Kingdom Research and Innovation (also known as UK Research and Innovation) [MC\_UP\_1201/33].
For the purpose of open access, the MRC Laboratory of Molecular Biology has applied a CC BY public copyright licence to any Author Accepted Manuscript version arising.
This project is supported through a research collaboration between AstraZeneca UK Limited and the Medical Research Council, reference BSF2-14.

\begin{footnotesize}
\bibliographystyle{ieeetr}
\bibliography{library}
\end{footnotesize}

\clearpage
\setcounter{page}{1}

\begin{center}
\section*{Training a force field for proteins and small molecules from scratch}
\section*{Supplementary Data}
\end{center}

\setcounter{figure}{0}
\setcounter{table}{0}
\renewcommand{\figurename}{Supplementary Figure}
\renewcommand{\tablename}{Supplementary Table}

\vspace{30px}

\begin{figure}[h]
  \centering
  \includegraphics[width=1.0\textwidth]{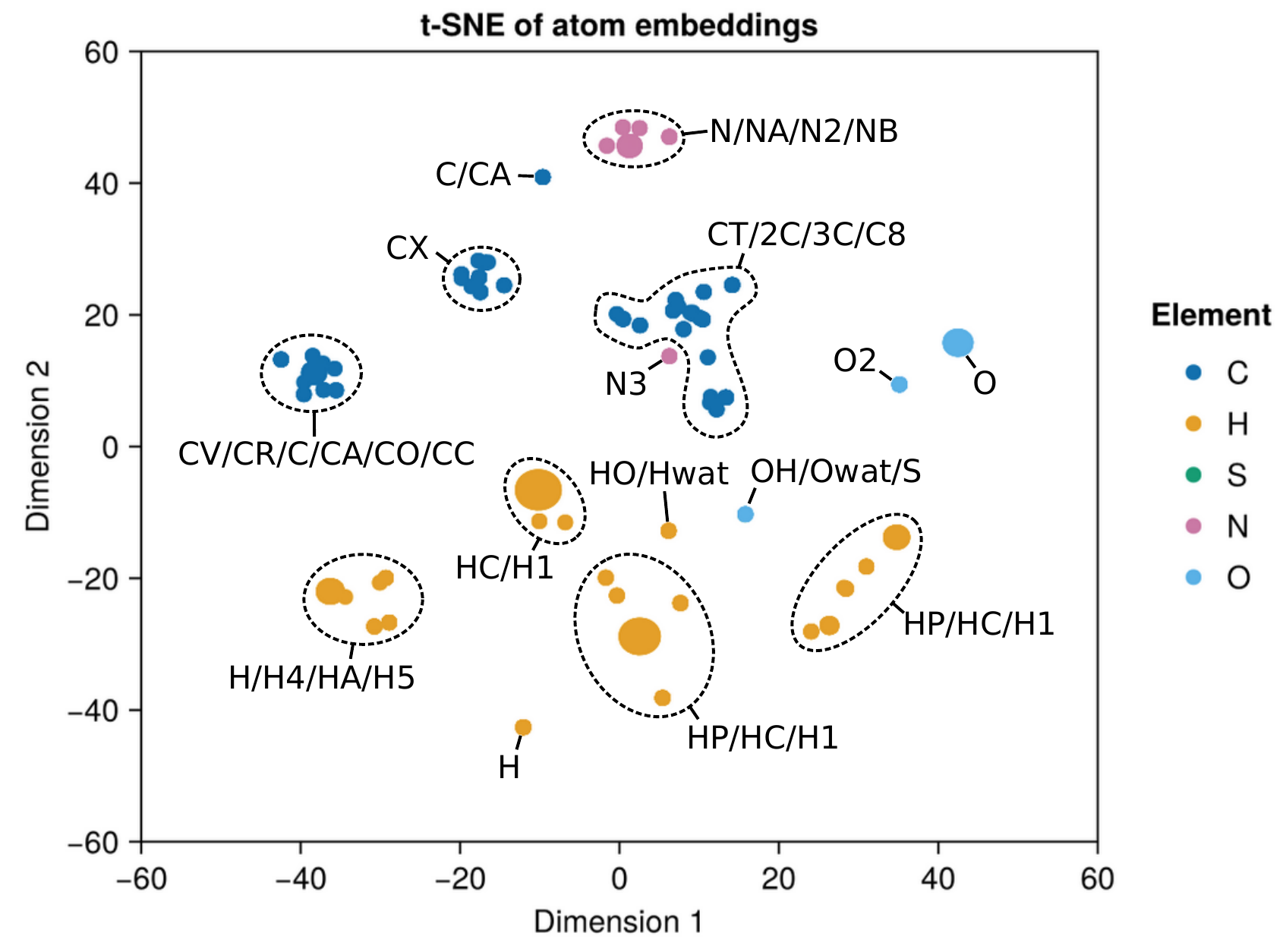}
  \caption{Exploring the embedding space of continuous atom types. Ubiquitin was parameterised with Garnet and the 64-dimensional embedding of each atom was calculated. t-SNE was used to reduce this to two dimensions for visualisation. The corresponding Amber14SB atom types found in different regions are shown, along with water atoms (Owat and Hwat).}
  \label{fig:embedding}
\end{figure}

\begin{figure}
  \centering
  \includegraphics[width=0.7\textwidth]{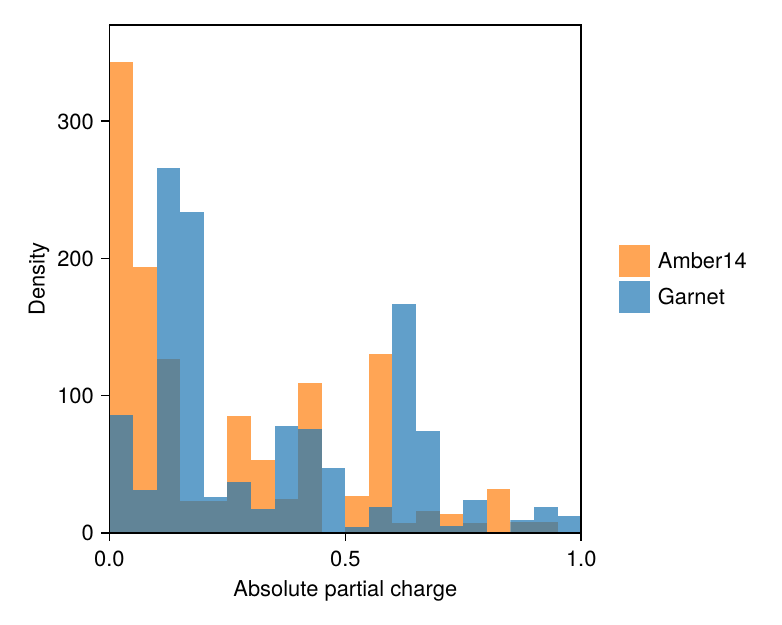}
  \caption{Partial charge distribution in Garnet. The partial charge of each atom in ubiquitin (1,231 atoms) is assigned with Garnet or Amber14SB \cite{Case2025}; the distribution of the absolute values of the partial charges is shown. For Garnet, the mean is 0.33 and the median is 0.19. For Amber14SB, the mean is 0.25 and the median is 0.11.}
  \label{fig:charges}
\end{figure}

\begin{figure}
  \centering
  \includegraphics[width=1.0\textwidth]{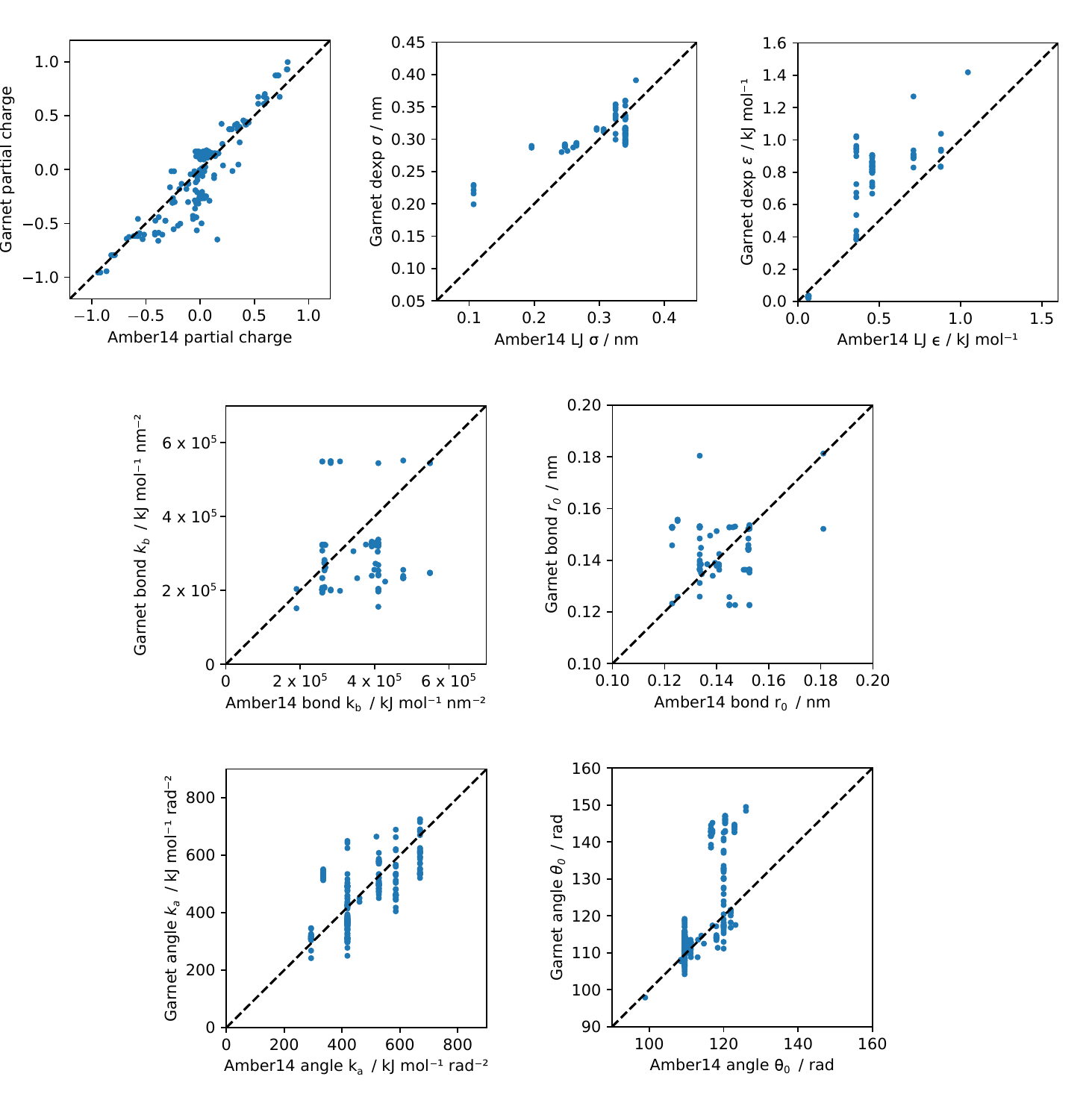}
  \caption{Comparison of parameters assigned to ubiquitin by Garnet and Amber14SB \cite{Case2025}. The functional forms of the Lennard-Jones and double exponential potentials are different, with $\sigma$ and $\varepsilon$ playing a similar role in both.}
  \label{fig:parameters}
\end{figure}

\begin{figure}
    \centering
    \includegraphics[width=0.95\linewidth]{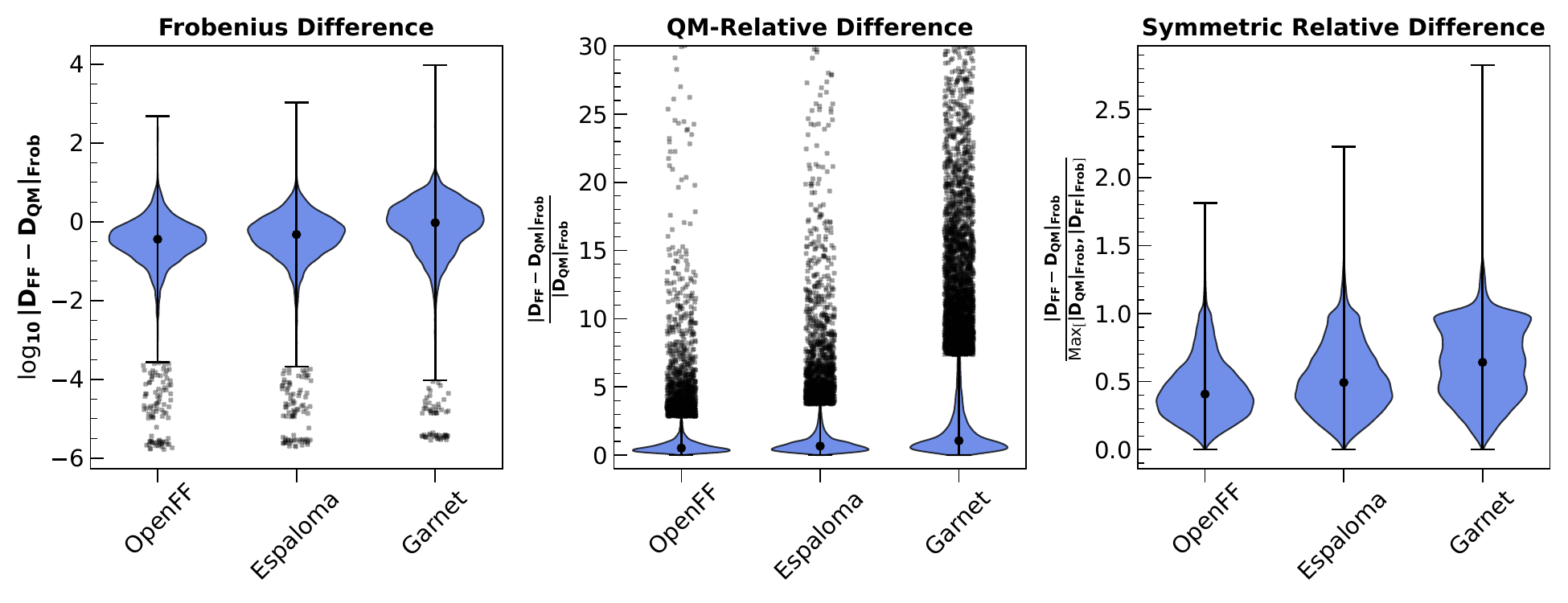}
    \caption{Comparison of dipole moments between OpenFF, Espaloma and Garnet, expressed as Frobenius norm differences of the dipole tensor.}
    \label{fig:dipole_tensor}
\end{figure}

\newpage
\FloatBarrier
\begin{table}[h!]
  \centering
  \footnotesize
  \setlength{\tabcolsep}{5pt}
  \renewcommand{\arraystretch}{1.15}
  \begin{tabular}{l r c c c c}
    \hline
    \textbf{Motif} & $\mathbf{N_\mathrm{all}}$ & \textbf{OpenFF} & \textbf{Espaloma} & \textbf{MACE-OFF} & \textbf{Garnet} \\
    \hline
    pyrazole & 11163 & -- & 1.41\,\,[1.22, 1.62] & 2.41\,\,[1.44, 3.71] & 1.27\,\,[1.10, 1.48] \\
    aldehyde & 78 & 8.24\,\,[3.69, 18.53] & 16.24\,\,[8.72, 30.43] & -- & 6.78\,\,[2.87, 16.30] \\
    thiol & 37 & 5.65\,\,[1.63, 21.25] & 5.65\,\,[1.63, 21.23] & -- & 5.65\,\,[1.63, 21.24] \\
    \hdashline
    pyrrole & 7533 & -- & 1.29\,\,[1.07, 1.55] & -- & 1.28\,\,[1.05, 1.54] \\
    phenol & 6044 & 2.47\,\,[2.13, 2.86] & -- & -- & 1.33\,\,[1.08, 1.63] \\
    carboxylic\_acid & 3160 & -- & 1.49\,\,[1.13, 1.96] & -- & 1.62\,\,[1.24, 2.11] \\
    indole & 3068 & -- & 1.71\,\,[1.31, 2.21] & -- & 1.74\,\,[1.34, 2.26] \\
    alkyne & 1339 & 1.88\,\,[1.27, 2.77] & 14.48\,\,[12.52, 16.69] & -- & -- \\
    diazo & 240 & 2.97\,\,[1.43, 6.17] & -- & 32.33\,\,[11.15, 88.06] & -- \\
    \hdashline
    fused\_aromatic & 10538 & 1.79\,\,[1.57, 2.02] & -- & -- & -- \\
    imidazole & 10014 & -- & -- & -- & 1.47\,\,[1.27, 1.70] \\
    fluorinated\_C & 8871 & -- & -- & -- & 1.31\,\,[1.10, 1.55] \\
    ketone & 3266 & -- & -- & -- & 1.95\,\,[1.53, 2.45] \\
    indazole & 2144 & -- & 1.59\,\,[1.14, 2.21] & -- & -- \\
    carboxylate & 1601 & 1.82\,\,[1.27, 2.61] & -- & -- & -- \\
    oxazole & 1346 & -- & 2.56\,\,[1.84, 3.56] & -- & -- \\
    imide & 969 & -- & -- & -- & 2.19\,\,[1.43, 3.32] \\
    azide & 16 & 32.96\,\,[11.78, 98.23] & -- & -- & -- \\
    \hline
  \end{tabular}
  \caption{Small molecule groups that appear more in the lowest 1\% of molecules by structure RMSD to QM structures after minimisation. The numbers represent the median and 95\% confidence bounds of the enrichment.}
  \label{tab:compare_structure_best}
\end{table}

\begin{table}[h!]
  \centering
  \footnotesize
  \setlength{\tabcolsep}{5pt}
  \renewcommand{\arraystretch}{1.15}
  \begin{tabular}{l r c c c c}
    \hline
    \textbf{Motif} & $\mathbf{N_\mathrm{all}}$ & \textbf{OpenFF} & \textbf{Espaloma} & \textbf{MACE-OFF} & \textbf{Garnet} \\
    \hline
    amide & 37001 & 1.30\,\,[1.22, 1.37] & 1.41\,\,[1.34, 1.48] & -- & 1.46\,\,[1.39, 1.53] \\
    tertiary\_amine & 21595 & 1.87\,\,[1.75, 2.00] & 1.34\,\,[1.23, 1.47] & -- & 1.38\,\,[1.26, 1.50] \\
    sulfonamide & 5463 & -- & 1.45\,\,[1.17, 1.78] & 2.67\,\,[1.59, 4.06] & 1.58\,\,[1.29, 1.93] \\
    quaternary\_amine & 4899 & 1.42\,\,[1.13, 1.77] & 1.73\,\,[1.41, 2.11] & -- & 1.64\,\,[1.33, 2.02] \\
    \hdashline
    sulfone & 9281 & -- & -- & 2.53\,\,[1.59, 3.70] & 1.29\,\,[1.09, 1.51] \\
    aniline & 6218 & 1.45\,\,[1.19, 1.76] & -- & -- & 1.29\,\,[1.05, 1.59] \\
    ester & 5876 & 1.69\,\,[1.41, 2.03] & 1.71\,\,[1.42, 2.05] & -- & -- \\
    disulfide & 57 & -- & 9.51\,\,[3.99, 23.03] & -- & 9.51\,\,[3.99, 23.02] \\
    \hdashline
    pyrimidine & 16159 & 1.31\,\,[1.16, 1.47] & -- & -- & -- \\
    ether & 13971 & -- & 1.41\,\,[1.24, 1.58] & -- & -- \\
    alkene & 7265 & -- & -- & -- & 1.87\,\,[1.60, 2.18] \\
    nitrile & 4298 & 1.35\,\,[1.05, 1.74] & -- & -- & -- \\
    urea & 3229 & -- & -- & -- & 1.94\,\,[1.52, 2.46] \\
    indazole & 2144 & -- & 2.27\,\,[1.72, 2.98] & -- & -- \\
    isoxazole & 1472 & -- & 1.71\,\,[1.16, 2.52] & -- & -- \\
    oxazole & 1347 & -- & -- & -- & 2.02\,\,[1.39, 2.93] \\
    conjugated & 722 & -- & -- & -- & 2.67\,\,[1.70, 4.15] \\
    sulfoxide & 588 & -- & -- & 2.37\,\,[1.48, 3.45] & -- \\
    guanidine & 436 & -- & 3.04\,\,[1.76, 5.23] & -- & -- \\
    enol & 222 & -- & 3.70\,\,[1.90, 7.29] & -- & -- \\
    phosphonate & 82 & -- & -- & -- & 6.43\,\,[2.76, 15.42] \\
    enoxide & 34 & -- & 6.18\,\,[1.81, 23.03] & -- & -- \\
    \hline
  \end{tabular}
  \caption{Small molecule groups that appear more in the highest 1\% of molecules by structure RMSD to QM structures after minimisation. The numbers represent the median and 95\% confidence bounds of the enrichment.}
  \label{tab:compare_structure_worst}
\end{table}

\FloatBarrier
\begin{table}
    \centering
    \begin{small}
    \begin{tabular}{lllccc}
        \hline
        \textbf{Property} & \textbf{Metric} & \textbf{Unit} & \textbf{GAFF} & \textbf{OpenFF-2.2.1} & \textbf{Garnet} \\
        \hline
        \multirow{3}{*}{$\Delta H_v$}
            & RMSE (Train) & kcal/mol & 1.14 & 0.36 & 0.13 \\
            & RMSE (Test)  & kcal/mol & 1.67 [0.56, 2.72] & 2.63 [0.48, 4.48] & 4.29 [2.12, 6.12] \\
            & Pearson's $r$ (Test) & - & 0.88 [0.74, 0.99] & 0.84 [0.69, 0.99]  & 0.81 [0.61, 0.97] \\
        \hline
        \multirow{3}{*}{$\rho$}
            & RMSE (Train) & g/mL & 0.045 & 0.017 & 0.045 \\
            & RMSE (Test) & g/mL & 0.021 [0.010,0.030] & 0.041 [0.021, 0.058] & 0.13 [0.11, 0.16] \\
            & Pearson's $r$ (Test) & - & 0.99 [0.99,0.99] & 0.99 [0.98, 0.99] & 0.93 [0.87, 0.99] \\
        \hline
    \end{tabular}
    \end{small}
    \caption{\revision{Performance of GAFF, OpenFF-2.2.1 and Garnet on predicting small molecule properties $\Delta H_v$ and $\rho$. RMSE and Pearson's $r$ between experimental and calculated $\Delta H_v$ and $\rho$ data are shown for the three force fields and separately for Garnet test and training data ($\Delta H_v$ values for three molecules were seen during training). The force fields were evaluated on the three training molecules as well as on 12 independent test molecules. Two test molecules had to be excluded from the GAFF evaluation, see the Methods. Numbers in brackets indicate bootstrapped 95\% confidence intervals. Since the training dataset consisted of only three molecules, bootstrapped confidence intervals and $r$ values are not reported for the training data.}}
    \label{tab:density_dhvap_results}
\end{table}

\begin{figure}
  \centering
  \includegraphics[width=1.0\textwidth]{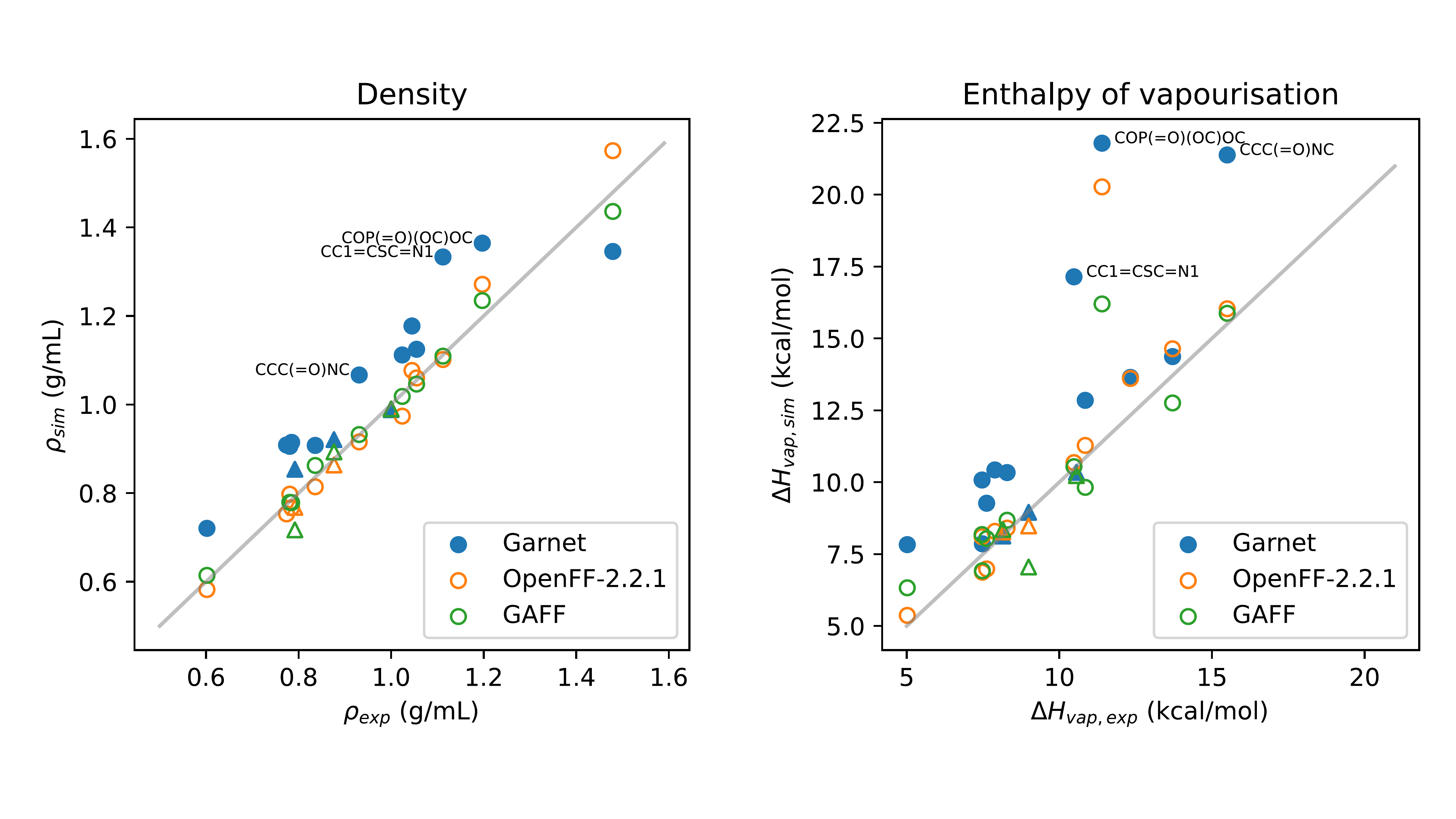}
  \caption{\revision{Performance of GAFF (green), OpenFF-2.2.1 (orange) and Garnet (blue) on predicting density ($\rho$) and enthalpy of vapourisation ($\Delta H_v$) for small molecules. Triangles indicate that $\Delta H_v$ data was used for Garnet training. SMILES strings are indicated for the three small molecules for which $\Delta H_v$ is considerably overestimated by Garnet.}}
  \label{fig:density_dhvap_results}
\end{figure}

\begin{figure}
  \centering
  \includegraphics[width=1.0\textwidth]{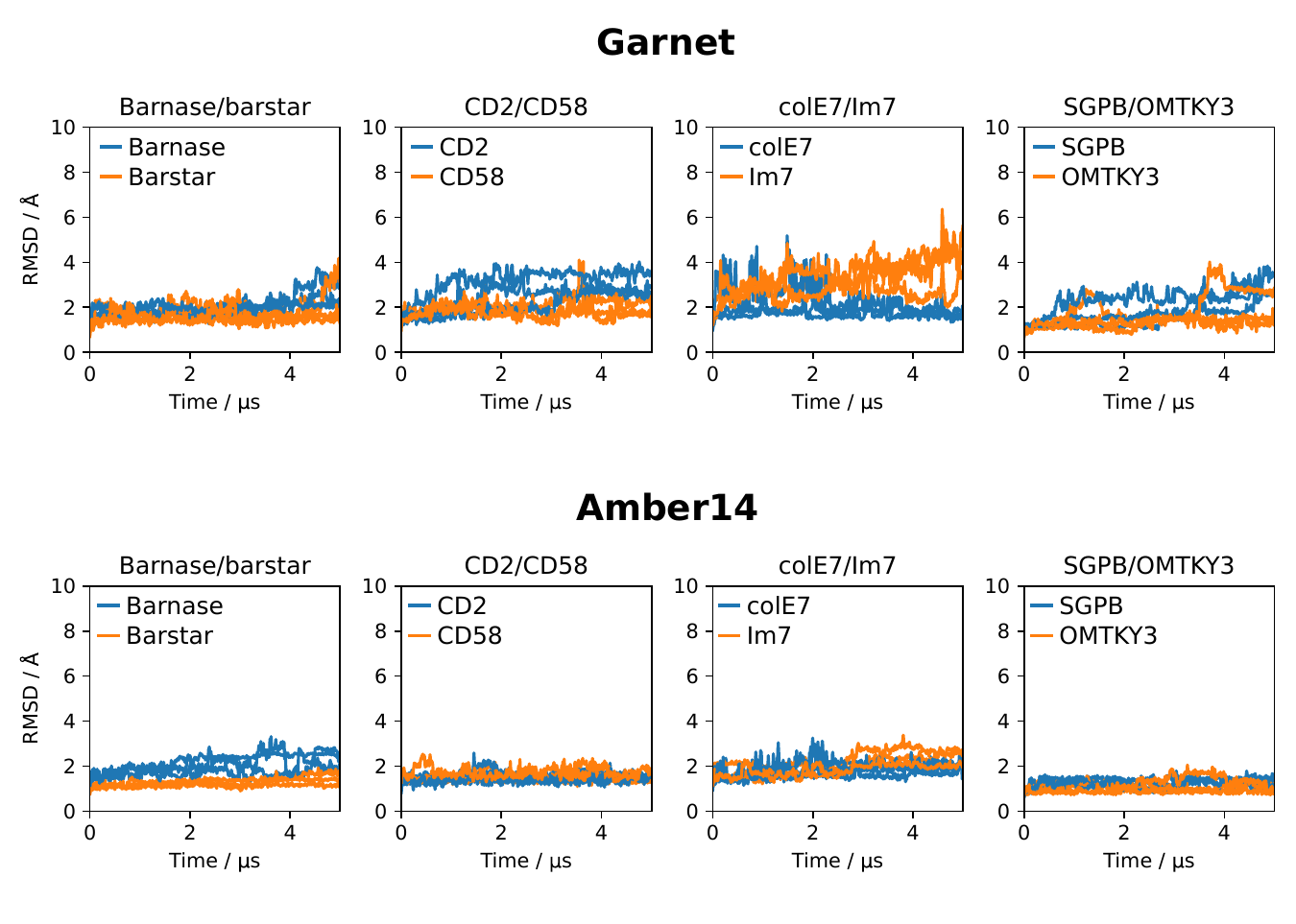}
  \caption{Performance of Garnet on simulating protein complexes. The simulations are the same as in Figure~\ref{fig:complex_idp}A. The RMSD of each monomer to its starting conformation is shown over the course of three repeats. The RMSD is smoothed by taking the mean over a window of values extending 10 snapshots either side.}
  \label{fig:complex_monomers}
\end{figure}

\begin{figure}
  \centering
  \includegraphics[width=1.0\textwidth]{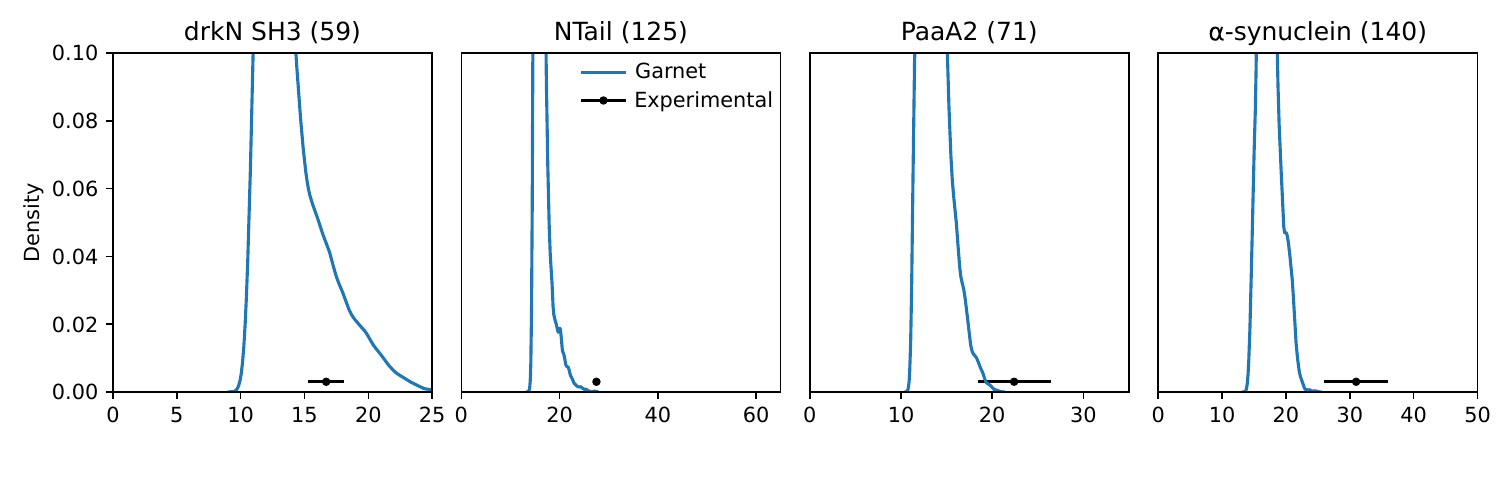}
  \caption{Radius of gyration when simulating IDPs with Garnet. The simulations are the same as in Figure~\ref{fig:complex_idp}B-C. The $R_g$ density is shown over the course of three repeats. The number in brackets is the number of residues in the protein.}
  \label{fig:idp_density}
\end{figure}

\begin{figure}
  \centering
  \includegraphics[width=1\textwidth]{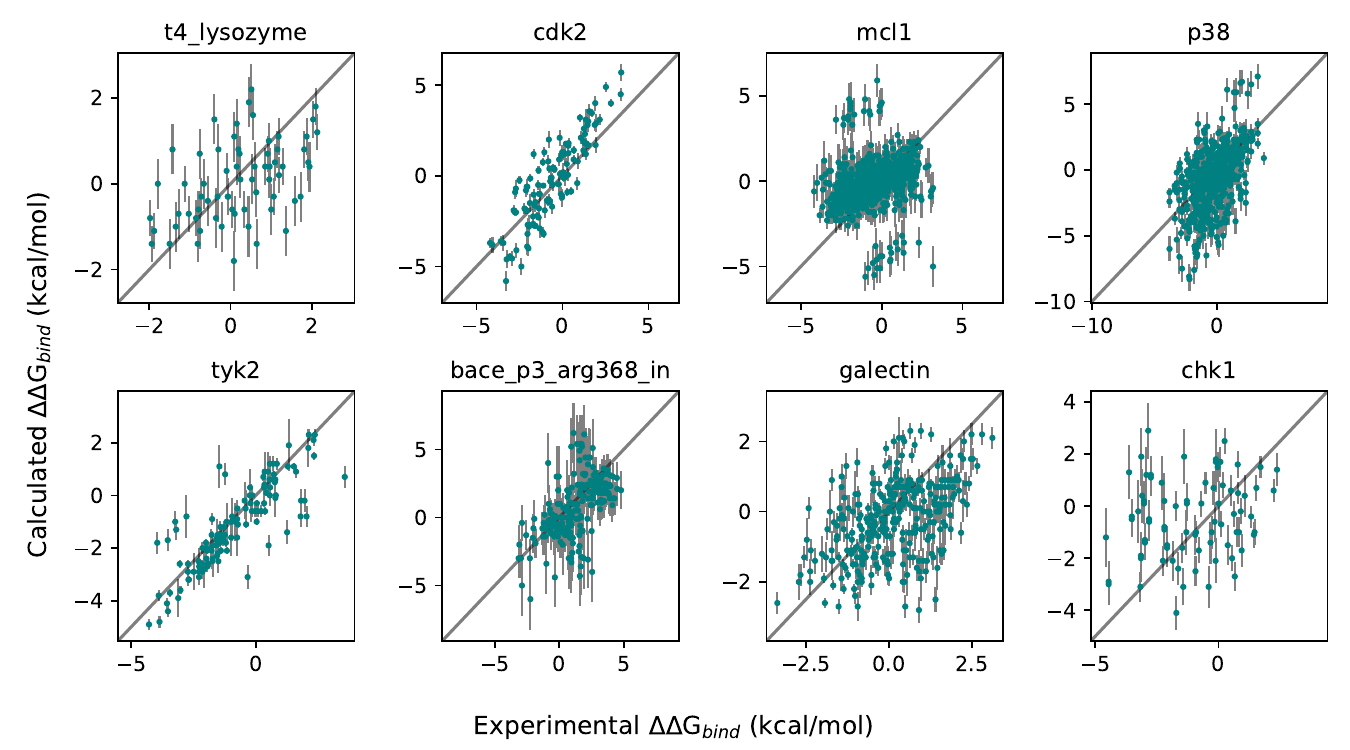}
  \caption{Garnet RBFE performance. Calculated and experimental $\Delta \Delta$G$_\mathrm{bind}$ values are shown for ligands binding to the eight proteins in our RBFE benchmark set. All (non-redundant) pairwise $\Delta \Delta$G$_\mathrm{bind}$ values are shown per ligand series (i.e.\ per protein). These data were used to estimate the RMSE values reported for Garnet in Figure~\ref{fig:rbfe}. Error bars are shown only for calculated $\Delta \Delta$G$_\mathrm{bind}$ and report uncertainties as calculated by OpenFE.}
  \label{fig:rbfe_scatter}
\end{figure}

\FloatBarrier
\begin{table}
    \centering
    \begin{small}
    \begin{tabular}{ l c c c }
        \hline
        \textbf{Force field} & \textbf{Pairwise RMSE (kcal/mol)} & \textbf{$\Delta$G Kendall $\tau$} & \textbf{Fraction of best ligands} \\
       \hline
       OpenFE   & 1.99 [1.82, 2.16] & 0.63 [0.38, 0.82]   & 0.55 \\
       FEP+     & 1.07	[1.00, 1.14] & 0.79 [0.66, 0.91]   & 0.74 \\
       Garnet   & 8.26 [7.74, 8.77] & -0.11 [-0.36, 0.17] & 0.16 \\
    \end{tabular}
    \end{small}
    \caption{RBFE results for protein cmet. A single ligand transformation in the ligand network for cmet involved a net charge change of 1, the effect of which was not well predicted by Garnet. The poorly-predicted edge singly connected all charged to all uncharged ligands in the transformation network. A poor $\Delta \Delta$G$_\mathrm{bind}$ estimate for the net charge change transformation caused $\Delta \Delta$G$_\mathrm{bind}$ estimates across the network to show a relatively large deviation from the experimental data, even though $\Delta \Delta$G$_\mathrm{bind}$ values between ligands of similar charges were relatively well-predicted (Supplementary Figure~\ref{fig:cmet}). Numbers in square brackets indicate bootstrapped 95\% confidence intervals.}
    \label{tab:cmet}
\end{table}

\begin{figure}
  \centering
  \includegraphics[width=0.45\textwidth]{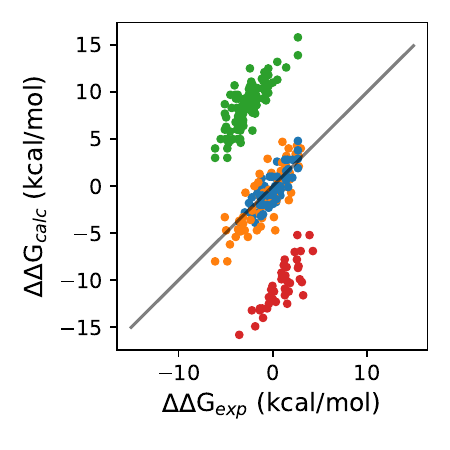}
  \caption{Garnet RBFE performance for the protein cmet. Calculated and experimental $\Delta \Delta$G$_\mathrm{bind}$ values are shown. Half of the cmet ligands were neutral, and the other half of the ligands had a net charge of +1. $\Delta \Delta$G values were relatively well-predicted (Kendall's $\tau$ = 0.63) for pairs of ligands with identical net charges (orange and blue data points). Absolute deviations to the experimental data were large for pairs with net charge differences (red and green data points), although correlations within each population of charge-changing pairs were relatively high (Kendall's $\tau$ = 0.60 for red cluster, Kendall's $\tau$ = 0.56 for green cluster). Ligand pairs are colour-coded using orange for neutral pairs, blue for pairs in which both ligands have a net charge of +1, green for pairs where the first ligand is neutral and the second charged, and red for pairs where the first ligand is charged and the second neutral.}
  \label{fig:cmet}
\end{figure}

\begin{figure}
  \centering
  \includegraphics[width=0.9\textwidth]{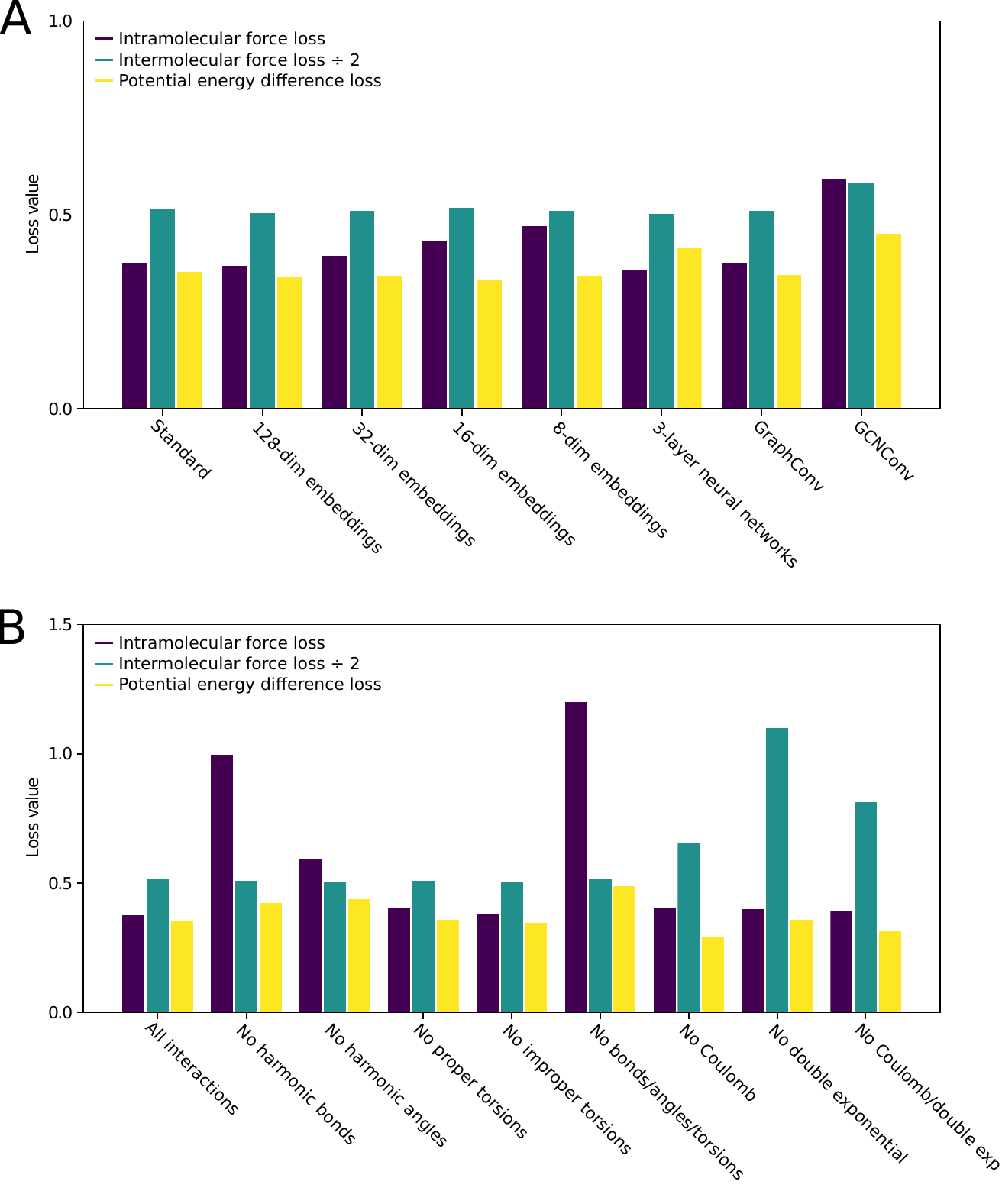}
  \caption{Assessing the performance of model variants and training models without certain interactions. For each variant 4 models were trained from scratch with no training simulations run, i.e.\ only the reference trajectories were used. This is different to the full training runs in Figure~\ref{fig:water}C. The lowest total loss of the 4 models after 7 epochs of training was calculated to select the best model shown here. For ease of visualisation, the intermolecular force loss is scaled. (A) Neural network hyperparameters. The standard case used 64-dimensional embeddings, 2-layer neural networks and the GraphSAGE convolution layer \cite{Hamilton2017}. (B) Removing components of the MM functional form. \revision{The double exponential potential is analogous to the Lennard-Jones potential and includes repulsive and attractive terms.}}
  \label{fig:ablations}
\end{figure}

\begin{table}
  \centering
  \begin{footnotesize}
    \begin{tabular}{ l r r r }
      \hline
      \textbf{Dataset} & \textbf{n confs} & \textbf{Weighting} & \textbf{n confs used} \\
      \hline
      SPICE Dipeptides Single Points Dataset v1.3           & 33,850    & 10  & 338,500   \\
      SPICE Solvated Amino Acids Single Points Dataset v1.1 & 1,300     & 100 & 130,000   \\
      SPICE DES370K Single Points Dataset v1.0 and v1.1     & 345,676   & 1   & 345,676   \\
      SPICE DES Monomers Single Points Dataset v1.1         & 18,700    & 1   & 18,700    \\
      SPICE PubChem Single Points Dataset v1.3              & 1,398,566 & 1   & 1,398,566 \\
      SPICE Solvated PubChem Set 1 v1.0                     & 13,934    & 20  & 278,680   \\
      SPICE Amino Acid Ligand v1.0                          & 194,174   & 2   & 388,348   \\
      SPICE Ion Pairs Single Points Dataset v1.2            & 1,426     & 10  & 14,260    \\
      SPICE Water Clusters v1.0                             & 1,000     & 100 & 100,000   \\
      SPICE PubChem Boron Silicon v1.0                      & 170,799   & 1   & 170,799   \\
      \hdashline
      RNA Single Point Dataset v1.0                         & 8,560     & 10  & 85,600    \\
      RNA Nucleoside Single Point Dataset v1.0              & 120       & 10  & 1,200     \\
      RNA Trinucleotide Single Point Dataset v1.0           & 6,080     & 10  & 60,800    \\
      \hdashline
      MACE-OFF water                                        & 1,681     & 100 & 168,100   \\
      \hdashline
      GEMS crambin                                          & 5,140     & 100 & 514,000   \\
      \hline
      Total                                                 & 2,201,006 & -   & 4,013,229 \\
      \hline
    \end{tabular}
  \end{footnotesize}
  \caption{DFT datasets used for training. The datasets are from SPICE \cite{Eastman2023, Eastman2024}, Espaloma \cite{Takaba2024}, MACE-OFF \cite{Kovacs2025} and GEMS \cite{Unke2024}. To ensure even training across all subsets, the data is weighted so that some subsets appear more often during training. Some of the data is retained as validation and test sets as described in the methods.}
  \label{tab:dataset}
\end{table}

\begin{figure}
  \centering
  \includegraphics[width=0.8\textwidth]{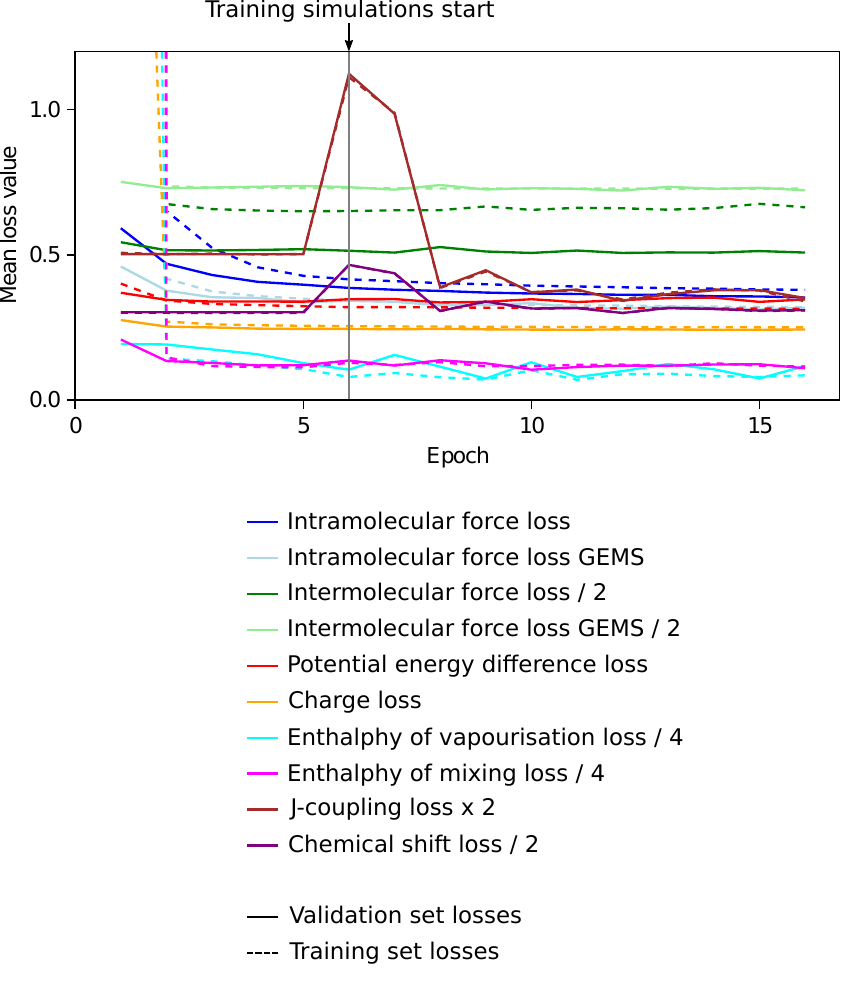}
  \caption{Garnet training progress. The various loss values are plotted over the course of training. Solid lines correspond to the validation set and dotted lines to the training set. Some losses are scaled as noted for ease of visualisation. Reference trajectories are used to train on condensed phase and protein properties before epoch 6. From epoch 6, simulations run with the training force field are used. The model after epoch 12 was treated as the final model and used for benchmarking.}
  \label{fig:training}
\end{figure}

\end{document}